\documentclass[10pt,final,conference]{IEEEtran}

\usepackage{balance}

\usepackage{siunitx}
\usepackage{graphicx}
\usepackage{amsmath} 
\usepackage{amssymb} 
\usepackage{amsthm} 
\usepackage[font=footnotesize,hypcap=true]{caption}
\usepackage[font=footnotesize,hypcap=true]{subcaption}

\usepackage{graphicx}
\usepackage[hyperref,x11names,cmyk,pdftex,table]{xcolor}

\usepackage{booktabs}
\usepackage{multirow}
\usepackage{dcolumn}

\usepackage{multicol} 

\usepackage[]{algorithm2e}

\usepackage{enumerate}
\usepackage{paralist}

\usepackage[resetfonts]{cmap}
\usepackage[T1]{fontenc}
\usepackage[utf8]{inputenc}

\usepackage[final,layout={inline},author=]{fixme}

\usepackage[pdftex,breaklinks=true,bookmarks=false]{hyperref}
\usepackage{cite}

\usepackage{cleveref}

\title{Using Graphlet Spectrograms for Temporal Pattern Analysis of
  Virus-Research Collaboration Networks}

\makeatletter
\let\titletext\@title
\makeatother

\author{
  \IEEEauthorblockN{%
    Dimitris~Floros\IEEEauthorrefmark{1}                     \qquad
    Tiancheng~Liu\IEEEauthorrefmark{2}        \qquad
    Nikos~Pitsianis\IEEEauthorrefmark{1}\IEEEauthorrefmark{2} \qquad
    Xiaobai~Sun\IEEEauthorrefmark{2}}
  \\
  \IEEEauthorblockA{\small%
    \begin{tabular}{c @{\qquad\qquad} c}
      \IEEEauthorrefmark{1}%
      Department~of~Electrical~and~Computer~Engineering
      &
      \IEEEauthorrefmark{2}%
      Department~of~Computer~Science
      \\
      Aristotle~University~of~Thessaloniki
      &
      Duke~University
      \\
      Thessaloniki~54124,~Greece
      &
      Durham,~NC~27708,~USA
    \end{tabular}%
  }
}

\date{} 

\newcommand{\pdfauthors}{%
  D. Floros, T. Liu, N. Pitsianis, X. Sun}

\graphicspath{%
  {./figures/}%
}

\hypersetup{%
  pdffitwindow=false,%
  pdfstartview={FitH},%
  colorlinks,%
  pdfauthor={\pdfauthors},%
  pdftitle={\titletext},%
  citecolor=PaleGreen4,%
  filecolor=DarkOrchid4,%
  linkcolor=OrangeRed4,%
  urlcolor=black%
}

\makeatletter
\def\th@plain{%
  \thm@notefont{}%
  \itshape %
}
\def\th@definition{%
  \thm@notefont{}%
  \normalfont %
}
\makeatother

\setkeys{Gin}{draft=false}

\pdfpageattr{/Group <</S /Transparency /I true /CS /DeviceRGB>>}

\let\leftorig\left
\let\rightorig\right
\renewcommand{\left}{\mathopen{}\mathclose\bgroup\leftorig}
\renewcommand{\right}{\aftergroup\egroup\rightorig}

\crefname{equation}{}{}

\crefname{figure}{Fig.}{Fig.}
\Crefname{figure}{Fig.}{Fig.}
\crefname{table}{Table}{Tables}
\Crefname{Table}{Table}{Tables}
\crefname{section}{Section}{Sections}
\Crefname{Section}{Section}{Sections}

\hypersetup{final}

\newcommand\blfootnote[1]{%
  \begingroup
  \renewcommand\thefootnote{}\footnote{#1}%
  \addtocounter{footnote}{-1}%
  \endgroup
}

\DeclareSIUnit{\nothing}{\relax}

\begin{document}

\bstctlcite{IEEEexample:BSTcontrol}

\maketitle

\addcontentsline{toc}{section}{Abstract}
\begin{abstract}
We introduce a new method for temporal pattern analysis of scientific
collaboration networks.  We investigate in particular virus research
activities through five epidemic or pandemic outbreaks in the recent
two decades and in the ongoing pandemic with COVID-19. Our method
embodies two innovative components.  The first is a simple model of
temporal collaboration networks with time segmented in publication
time and convolved in citation history, to effectively capture and
accommodate collaboration activities at mixed time scales.
The second component is the novel use of graphlets to encode
topological structures and to detect change and persistence in
collaboration activities over time. We discover in particular two
unique and universal roles of bi-fork graphlet in (1) identifying
bridges among triangle clusters and (2) quantifying grassroots as the
backbone of every collaboration network.
We present a number of intriguing patterns and findings about the
virus-research activities.

\blfootnote{\vspace*{-0.7em}The first two authors contributed
   equally to this work.\vspace*{-1em}}

\end{abstract}


\section{Introduction}
\label{sec:introduction}

\begin{figure}
  \centering
  \begin{subfigure}{.48\linewidth}
    \centering
    \includegraphics[height=0.115\textheight]{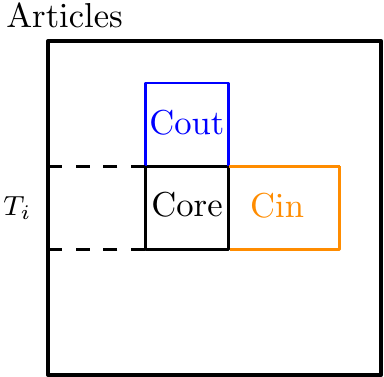}
  \end{subfigure}
  \begin{subfigure}{.48\linewidth}
    \centering
    \includegraphics[height=0.10\textheight]{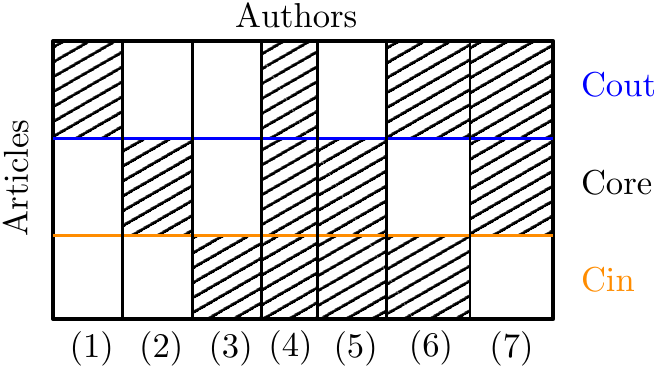}
  \end{subfigure}
  \\[0.5em]
  \begin{subfigure}{.40\linewidth}
    \centering
    \includegraphics[height=0.11\textheight]{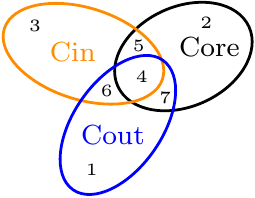}
  \end{subfigure}
  \begin{subfigure}{.48\linewidth}
    \centering
    \includegraphics[height=0.15\textheight]{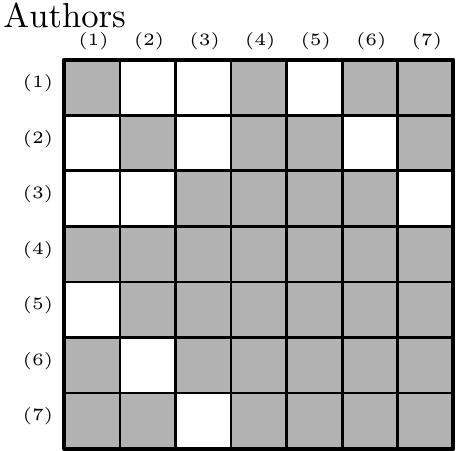}
  \end{subfigure}
  \\[0.5em]
  \caption{Schema for extracting epoch-centered citation network and
    time-segmented collaboration network. In left right and top down order: 
    {\bf (a)} Epoch-centered triad (open)
    citation network  extracted from the universal literature graph, which is
    represented by the upper triangular adjacency matrix, with article
    nodes sorted in chronological order, the first row identifies with
    the earliest article. The triad open network centers at the block
    \textcolor{black}{\bf Core} on the
    diagonal, which represents the subgraph within time window or
    epoch $T_i$. It also includes two open components, block
    \textcolor{DeepSkyBlue2}{\bf Cout} composed of outward links in
    one hop to precursor articles and block \textcolor{orange}{\bf
      Cin} composed of inward links in one hop from
    follower articles.
    {\bf (b)} The article-author bipartite with the articles
    partitioned into {\tt Cout}, {\tt Core} and {\tt Cin} subsets and
    the involved authors grouped into seven cohorts by the
    non-overlapping septa-partition schema shown next.
    {\bf (c)} The septa-partition, shown in the Venn diagram, of the authors
    involved in the triad citation network.
    {\bf (d)} The time-segmented author collaboration network
    represented by a symmetric adjacency matrix in $7 \!\times\! 7$
    block partition. There are $6$ empty blocks in the upper/lower
    triangular. }
  \label{fig:pipeline}
\end{figure}

\begin{figure}[!hb]
  \vspace*{-1em}
  \centering
  \newcommand{\figGraphletWidth}{0.14\linewidth}
  \newcommand{\graphletID}{0}
  \newcommand{\graphletName}{0}
  \hspace*{\fill}
  \renewcommand{\graphletID}{0}
  \renewcommand{\graphletName}{0}
\begin{subfigure}{\figGraphletWidth}
  \includegraphics[width=\linewidth]{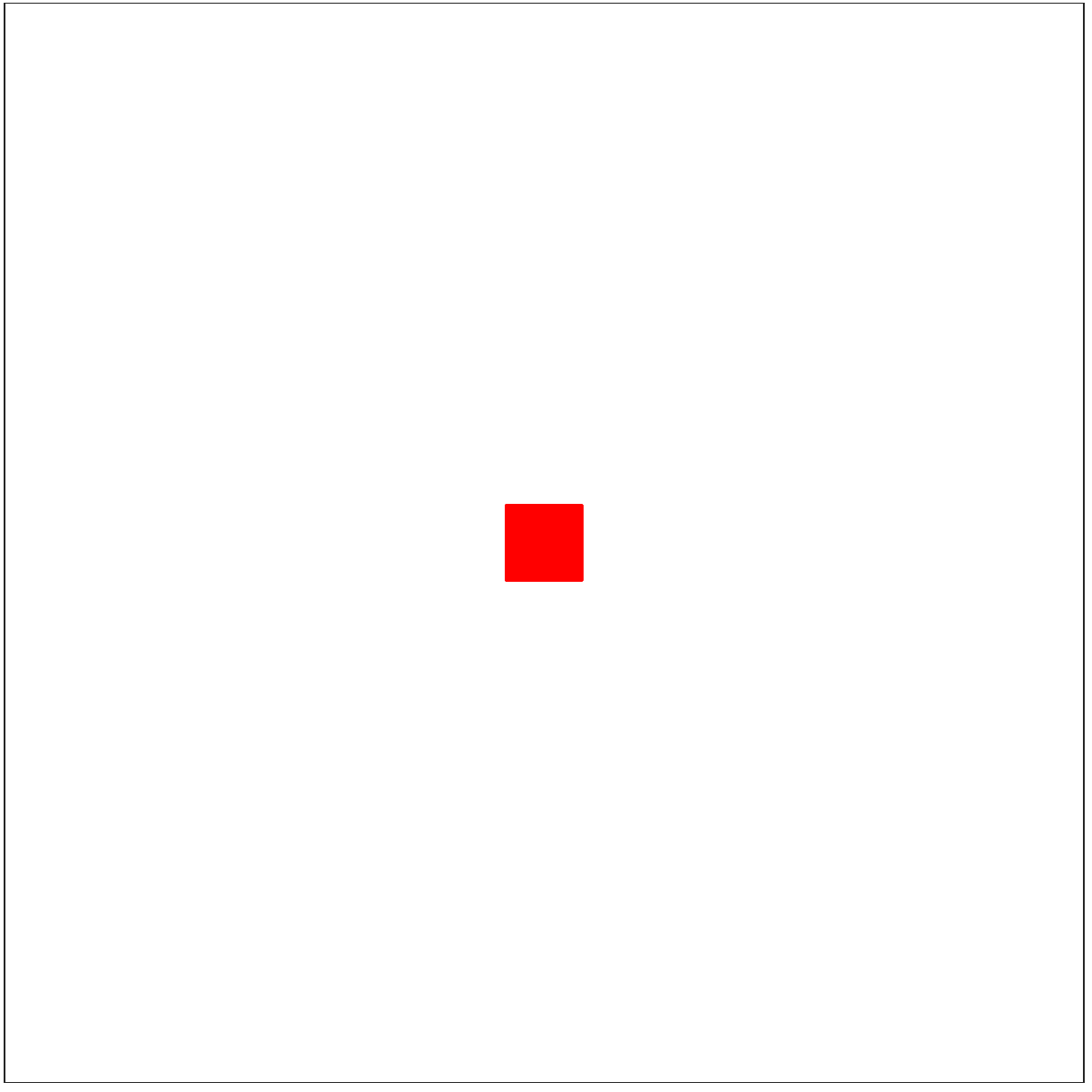}
  \vspace*{-1.5em}
  \caption*{$\sigma_{\graphletName}$}
\end{subfigure}
\hspace*{\fill}
  \renewcommand{\graphletID}{1}
  \renewcommand{\graphletName}{1}
\begin{subfigure}{\figGraphletWidth}
  \includegraphics[width=\linewidth]{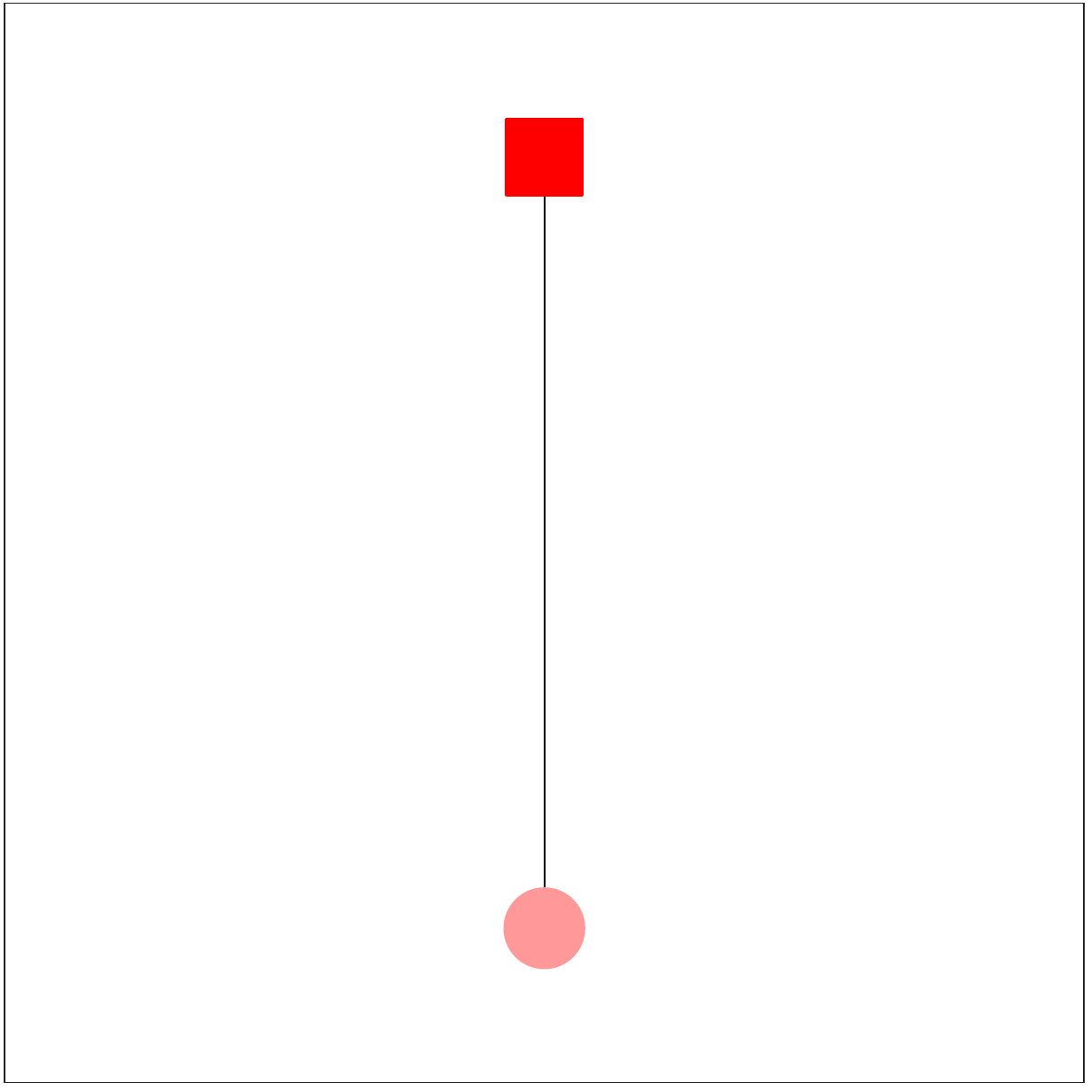}
  \vspace*{-1.5em}
  \caption*{$\sigma_{\graphletName}$}
\end{subfigure}
\hspace*{\fill}
  \renewcommand{\graphletID}{2}
  \renewcommand{\graphletName}{2}
\begin{subfigure}{\figGraphletWidth}
  \includegraphics[width=\linewidth]{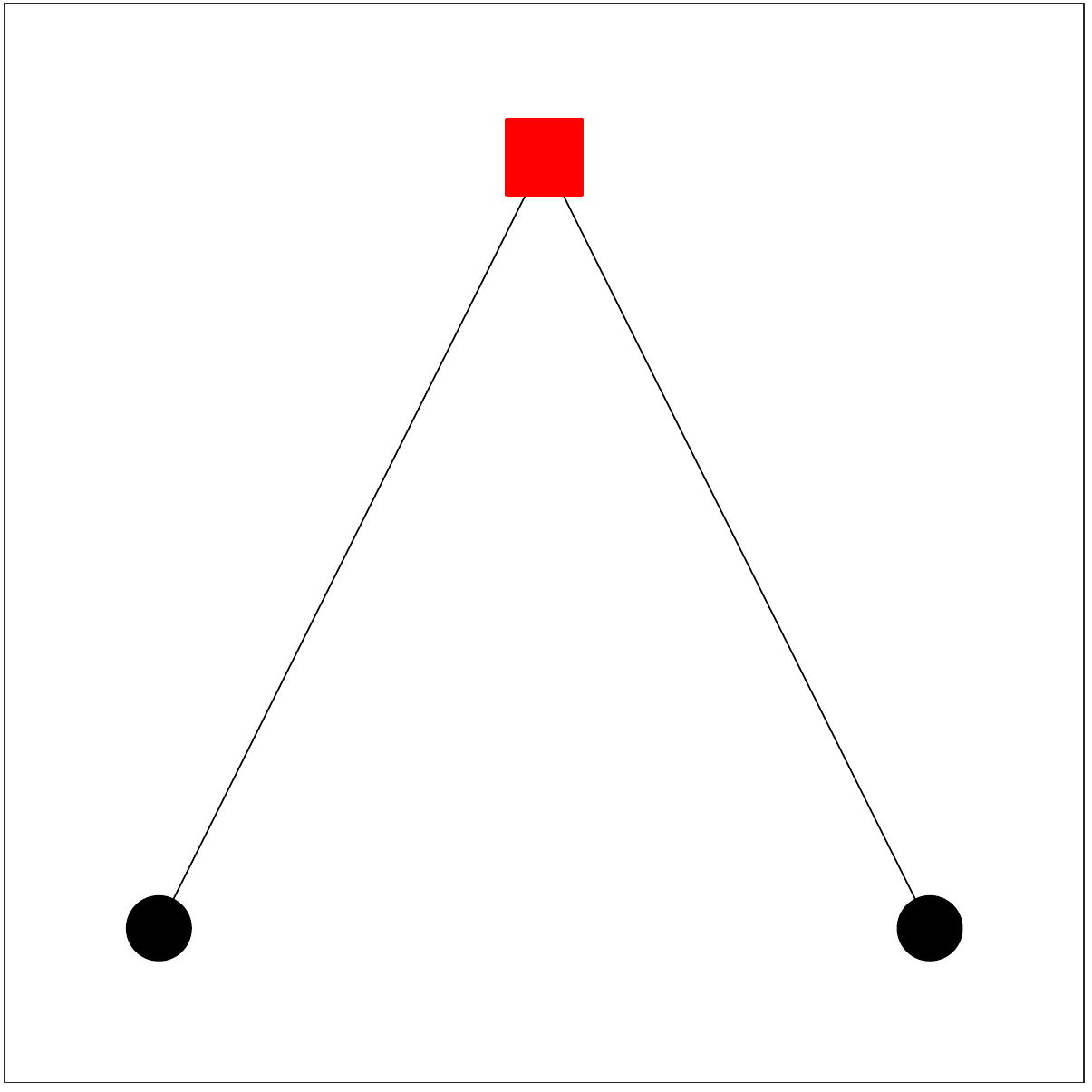}
  \vspace*{-1.5em}
  \caption*{$\sigma_{\graphletName}$}
\end{subfigure}
\hspace*{\fill}
  \renewcommand{\graphletID}{3}
  \renewcommand{\graphletName}{3}
\begin{subfigure}{\figGraphletWidth}
  \includegraphics[width=\linewidth]{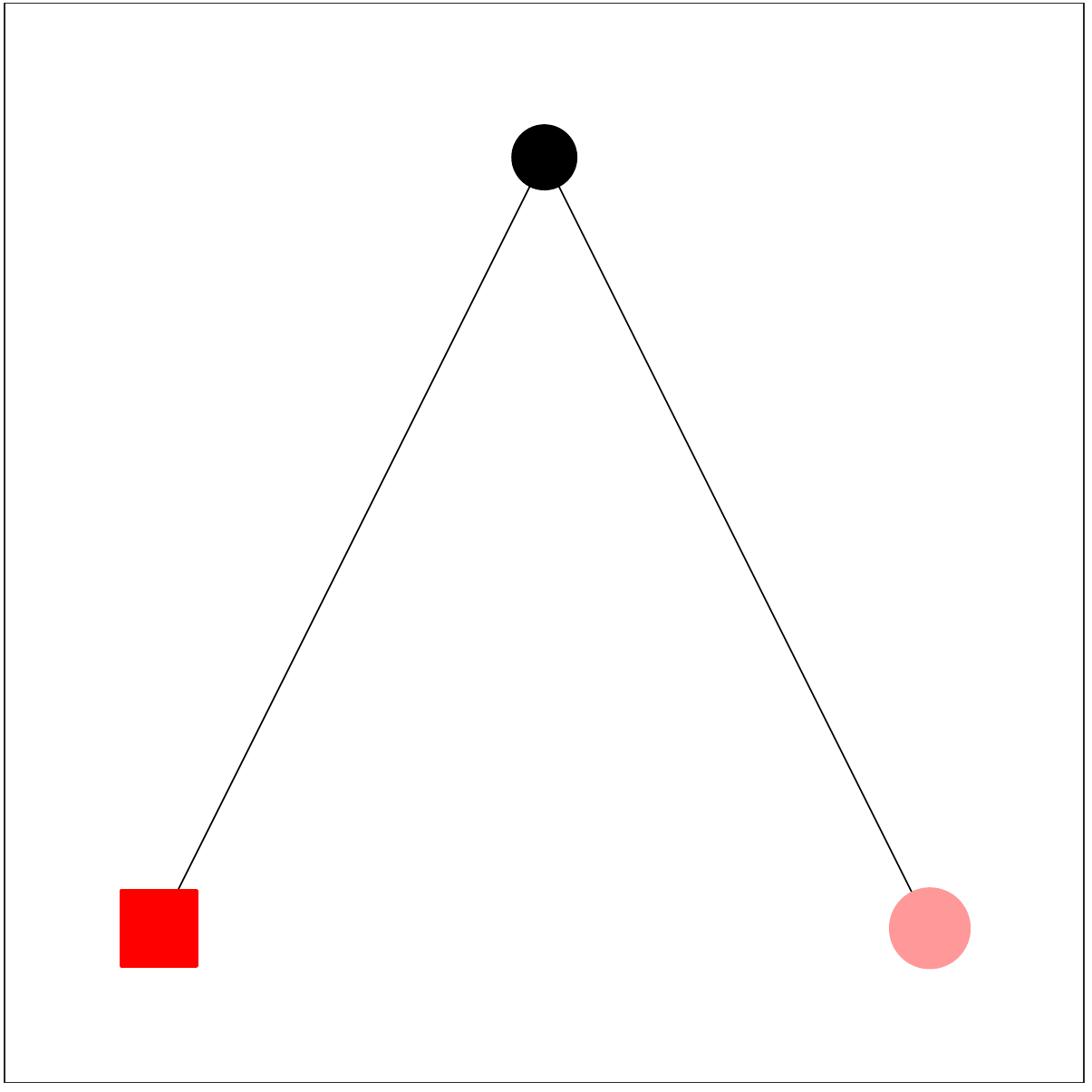}
  \vspace*{-1.5em}
  \caption*{$\sigma_{\graphletName}$}
\end{subfigure}
\hspace*{\fill}
  \renewcommand{\graphletID}{4}
  \renewcommand{\graphletName}{4}
\begin{subfigure}{\figGraphletWidth}
  \includegraphics[width=\linewidth]{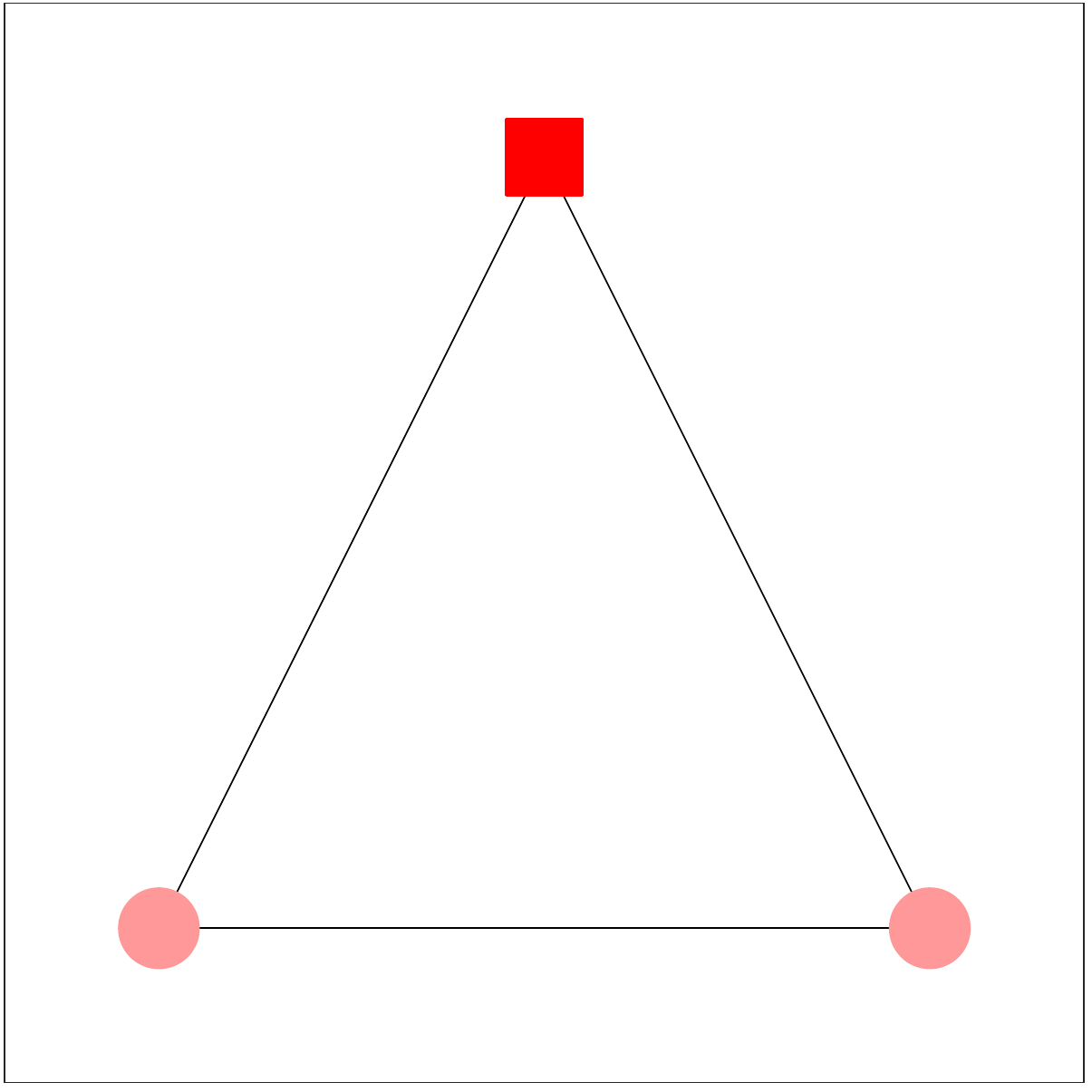}
  \vspace*{-1.5em}
  \caption*{$\sigma_{\graphletName}$}
\end{subfigure}
\hspace*{\fill}
  \\[1em]
  \caption{A dictionary of 5 graphlets with 1 to 3 nodes. There are
    the singleton $\sigma_0$, edge $\sigma_1$, bi-fork $\sigma_2$
    ($K_{1,2}$), 2-path $\sigma_3$ ($P_2$) and triangle $\sigma_4$
    ($C_3$ a.k.a. $K_3$).  In each graphlet, the red square node
    specifies the designated incidence node, nodes in light red are
    automorphic to the incidence node.
  }
  \label{fig:graphlets}
\end{figure}

We present a new approach for uncovering and analyzing temporal
patterns of author collaboration networks over different time
periods. Author collaboration networks are among the most studied over
a half century \cite{goffman1969,odda1979,newman2001, molontay2020}.
We make three key contributions.
First, we introduce a live literature graph (LG) we created in March
2020~\cite{floros2020a} and have made frequent updates since. The
literature body is mainly on research of various viruses, including
HIV, SARS, Swine flu, MERS, Ebola, Avian flu and the coronavirus
responsible for the ongoing, ravaging pandemic with COVID-19. The LG
contains articles that date back to 1744.
For temporal pattern detection and analysis, we extract $6$ open
citation networks over different time windows, $6$ related
article-author bipartite graphs and $6$ author collaboration
networks. The basic information is in \Cref{tab:epochs}. We apply our
temporal analysis approach to these networks and use them in turn as
real-world network examples. We describe the LG in \Cref{sec:LG-covid19-hotp}.
Secondly we introduce an original method for constructing author
collaboration networks with time segmented in publication time and
convolved (overlapped) in citation history.  This is a significant
deviation from conventional methods, in which temporal changes in data
was neglected in static analysis, or simplistically sliced by time
windows at a macroscopic scale, or regulated by time-dependent models
at a microscopic time scale.  We describe our model of time-shifted
and time-convolved networks in \Cref{sec:triad-subgraphs}.

\begin{table}[bht]
  \centering
  \caption{Time windows and article-author bipartite graphs associated
    with $6$ epoch-centered triad citation networks and the entire
    \texttt{LG-covid-HOTP} up to the present. See also (a) and (b) of
    \Cref{fig:pipeline}.}
  \resizebox{0.95 \linewidth}{!}{%
    \begin{tabular}{lcrrrcc}
      \toprule
      Epoch & Period ($T$) & \#Articles & \#Authors & \#Links &
                     $\frac{\mbox{Authors}}{\mbox{Article}}$  &
                     $\frac{\mbox{Articles}}{\mbox{Author}}$ \\ 
      \midrule
      SARS & 2002-2004 & \num{52374} & \num{181716} & \num{326064} &
                                                                     \num{6.2} & \num{1.8} \\
      Swine flu & 2009-2011 & \num{78974} & \num{266818} & \num{519529} &
                                                                          \num{6.6} & \num{1.9} \\
      MERS & 2012-2014 & \num{84810} & \num{293021} & \num{577120} & \num{6.8} &
                                                                           \num{2.0} \\
      Ebola & 2014-2016 & \num{85878} & \num{305842} & \num{599450} & \num{7.0} &
                                                                            \num{2.0} \\
      Avian flu & 2017-2019 & \num{79849} & \num{313256} & \num{587226} &
                                                                          \num{7.4} & \num{1.9} \\
      COVID-19 & 2020-\phantom{2020} & \num{21664} & \num{113357} &
                                                                    \num{166907}
                                                           & \num{7.7} & \num{1.5}
      \\
      \midrule
      LG-covid19 & 1744-\phantom{2020} & \num{251551} & \num{551713} & \num{1086779} & \num{4.3}  & \num{2.0}\\
      \bottomrule
    \end{tabular}%
  }
  \vspace*{-15pt}
  \label{tab:epochs}
\end{table} %

\begin{figure}
  \centering
  \vspace*{0.5em}
  \includegraphics[width=\linewidth]{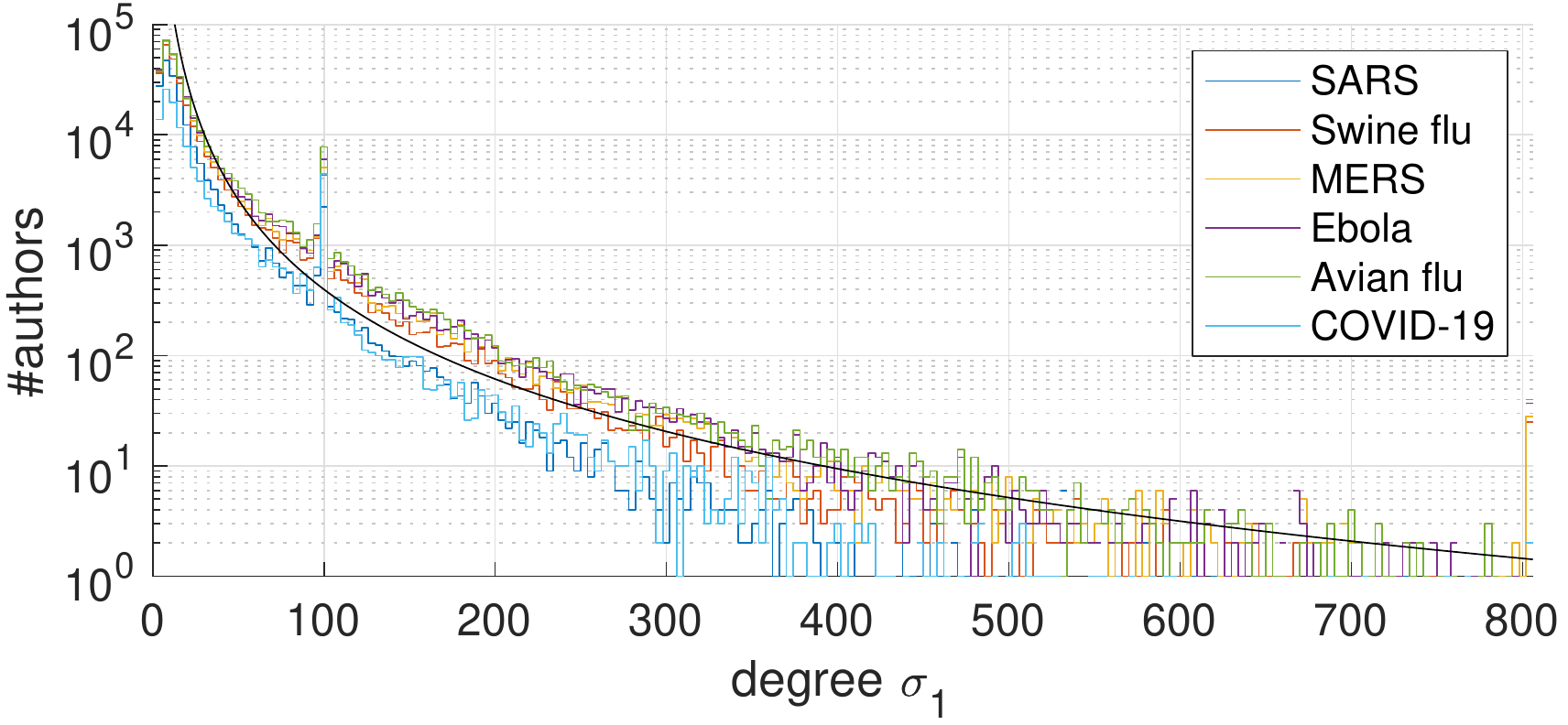}
  \vspace*{-1.5em}
  \caption{Degree ($\sigma_1$) histograms for $6$ collaboration
    networks associated with the epochs specified in
    \Cref{tab:epochs}.  The bin size is \num{4} except the last one
    that includes all authors with degree above $800$.  %
    The network associated with COVID-19 is over a time window of only
    $6$ months up to the present, 5 times as small as the 3-year
    periods  for the others.  %
    The burst at 100, common to all $6$ histograms, is an artifact caused by
    the cutoff threshold at $100$ set by \text{Scopus} (and similarly
    by other data sources) over the number of authors per
    article. The artifact is also manifested in \cref{fig:sigma-curves} and
    explained in detail there. 
    {\bf Observation}.  All six degree distributions follow the
    power-law pattern.
   A synthetic power law curve with parameter $\gamma = 2.7$
    is provided and shown in black to serve as a reference. 
   }
  \label{fig:degree-histogram-sigma1}
\end{figure}

Thirdly, we use graphlets as coding elements for encoding topological
structures and dynamic changes of live networks.
In \Cref{fig:graphlets} the first five generic graphlets are shown.
Graphlets are fundamental topology elements of all networks or graphs.
In concept, graphlets for network analysis are similar to wavelets for
spectro-temporal analysis of signal processing~\cite{rioul1991},
shapelets for time series classification~\cite{ye2009}, super-pixels
for image analysis~\cite{ren2003}, and n-grams for natural language
processing~\cite{shannon1948,shannon1951}.
Network analysis using graphlets has advanced in recent years, since
the original work by Pr\v{z}ulj {\em et al} in 2004
\cite{przulj2004,yaveroglu2015,sarajlic2016,shervashidze2009}.
Graphlets are mostly used for statistical characterization and
modeling of entire networks.
We recently established a new way of using graphlets to encode
microscopic structure at vertex neighborhood to macroscopic structure
such as cluster configurations~\cite{floros2020}.
In this work, we succeed in using graphlet encoding schemes to detect
changes and persistence across author collaboration networks at
different epochs, while the conventional approach based on degree
distributions is short of such differentiation capacity, as shown in
\Cref{fig:degree-histogram-sigma1}.

We present in \Cref{sec:temporal-pattern-analysis} a few remarkable
findings.

\section{Live literature graph: LG-covid19-HOTP}
\label{sec:LG-covid19-hotp}

We introduce briefly a particular real-world literature graph, {\tt
  LG-covid19-HOTP}~\cite{floros2020a}, created by three of the authors in early
  March, released onto the Internet on March 26, 2020, and updated
  weekly ever since.\footnote{
  \url{https://lg-covid-19-hotp.cs.duke.edu}; data snapshot is taken
  in June, 2020%
  }
The literature body, centering on virus research, is composed of
COVID-19 scholarly articles, their precursor and contemporary studies
on viruses.
Our data collection starts with a set of seed articles (not a single
ego article) and makes backward (citing) and forward (being cited)
spans from several very large literature
databases~\cite{ammar2018}.  In principle, the collection is by
preferential attachment~\cite{barabasi2016}.
Detailed collection information can be found at the home website.
There are several other COVID-19 themed literature
datasets\footnote{\url{https://www.kaggle.com/search?q=COVID-19}\vspace*{-1em}},
some of them are absent of citation links.

\begin{figure}
  \centering
  \includegraphics[width=\linewidth]{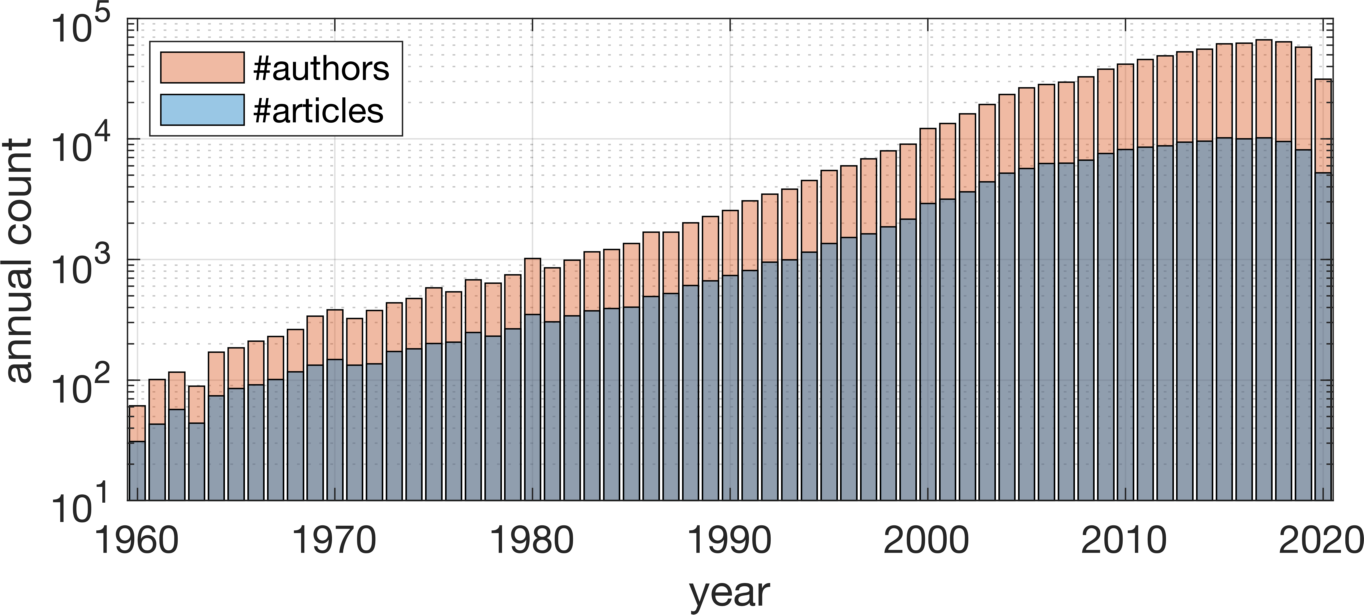}
  \caption{Histograms of annual counts (in logarithmic scale)
    for published articles  (in blue) and the
    authors (in red) between 1960 and  2020-June, by the
    records in {\tt LG-covid19-HOTP}. The total number of authors
    is $551,713$, more than twice the total number of articles,
    $251,551$.
    {\bf Observation:} Both annual counts increase steadily over time
    (the lower dips at 2018-19 are mostly due to latency in data
    collection and registration).  A burst is expected by the end of
    2020 given the current counts in less than 6 months. }
  \label{fig:author-article-histogram}
\end{figure}

We provide in \Cref{fig:author-article-histogram} and
\Cref{tab:epochs} the basic quantitative information of {\tt
LG-covid19-HOTP}. The number of articles and the author population
increase steadily, without noticeable bursts. The dips in recent two
years are due to update latency in the databases our collection relies
on.  We expect, however, a burst in 2020 by the current collection up
to June.

Our temporal pattern analysis rests on recognizing and exploring
important properties of the literature graph.  %
{\tt LG-covid19-HOTP} is of multiple attribute dimensions, or a
multiplex network. It has several types of vertices/nodes and several
types of edges/links.  The primary nodes are articles;
the primary edges are citation links from citing articles to cited articles.
The adjacency matrix for the citation network among the
article nodes is nearly upper triangular, where the articles are ordered
chronologically, the first row identifies with the earliest
article. We depict the citation adjacency matrix at the top-left of
\Cref{fig:pipeline}.
Although primary, the article nodes are actually the results of
actions by the author nodes. The LG contains the bipartite between
article nodes and author nodes. An article is written by one or more
authors; and an author is connected to one or more 
articles.  We depict the bipartite incidence matrix at the
top-right of \Cref{fig:pipeline}.
Via the bipartite, we get author collaboration network/adjacency
matrix at the bottom-right of \Cref{fig:pipeline}.
Other nodes represent author affiliations, author profiles, 
semantic entities in titles, abstracts and text bodies, and
data figures and tables. 

More importantly, the LG is a live network, changing and evolving
incessantly, with growth and collectively selected memory, not
ephemeral nor static.  However, author collaboration networks had been
largely made static (by time integration).
The publication timestamp with each article records the birth time of
the article, which may herald a new path or trajectory. The
bibliographical references are links to selected precursor work in
time as well as in concept.  Some other earlier work may be forgotten
for a while or for good, not as indelible as seemed in data records.
We investigate on temporal differentiation and persistence across
author collaboration networks at different epochs.

\begin{figure}
  \centering
  \begin{subfigure}{0.48\linewidth}
    \centering
    \includegraphics[width=.8\linewidth]{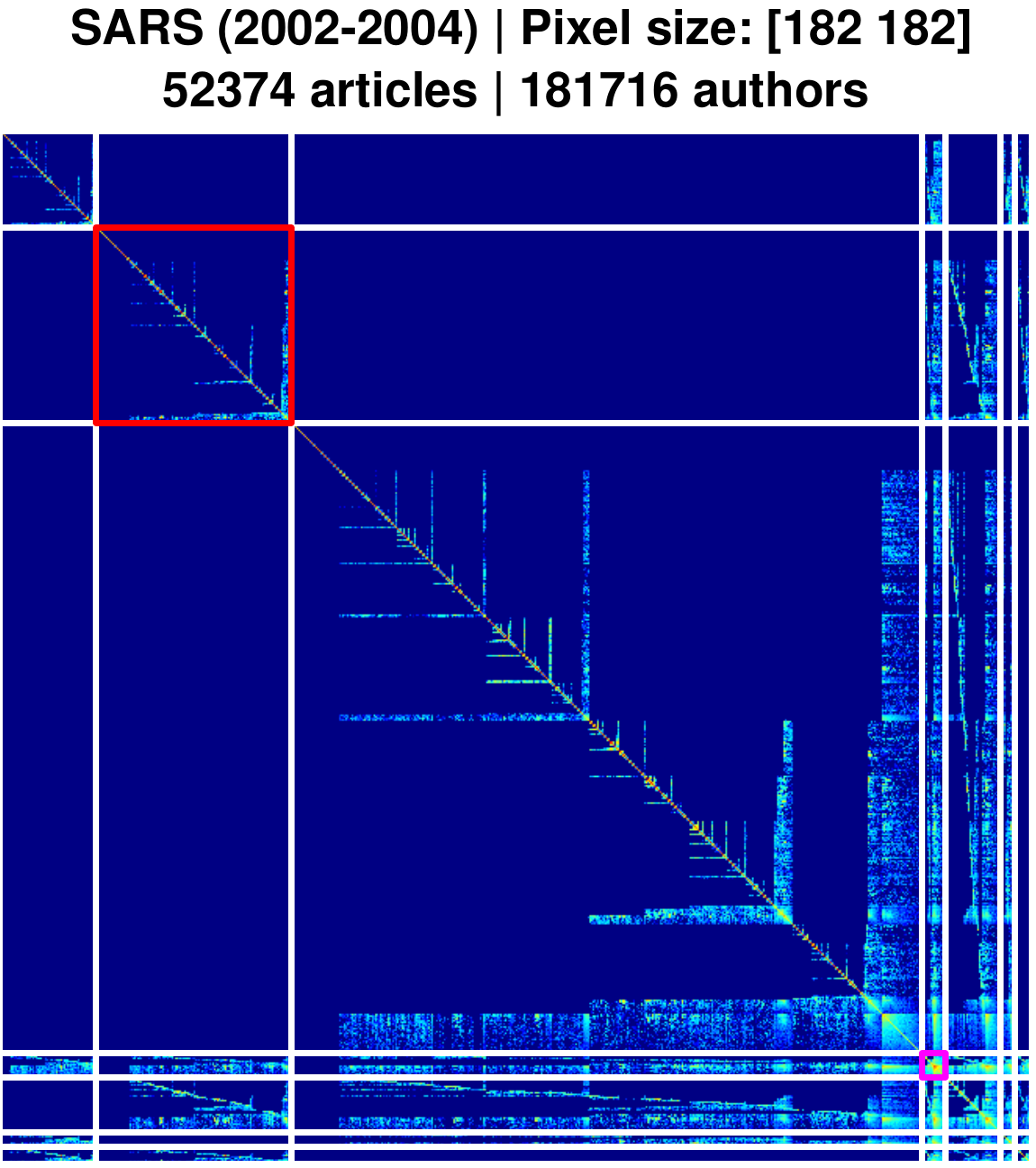}
  \end{subfigure}
  \begin{subfigure}{0.48\linewidth}
    \centering
    \includegraphics[width=.8\linewidth]{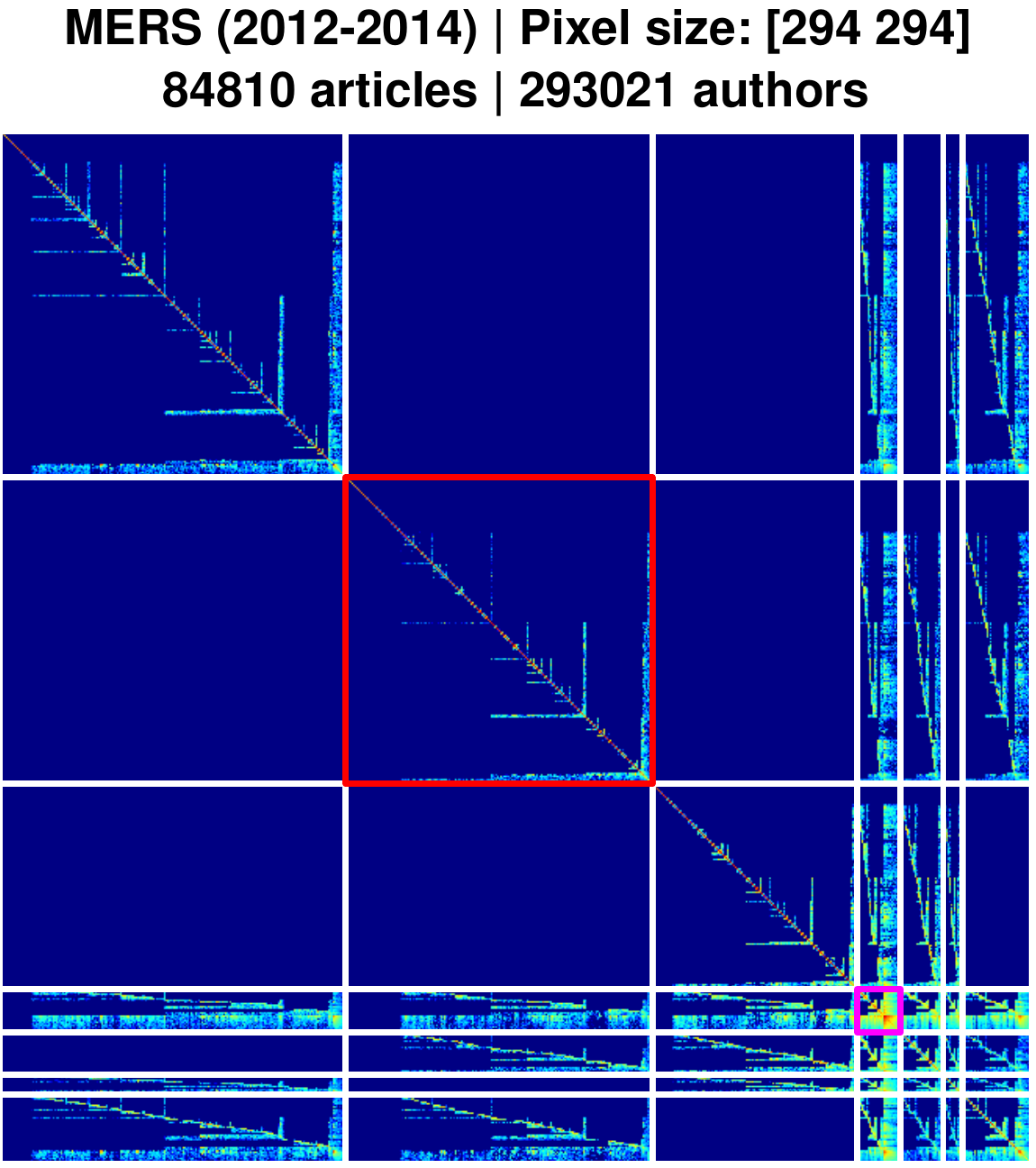}
  \end{subfigure}
  \\  [0.5em]
  \begin{subfigure}{0.48\linewidth}
    \centering
    \includegraphics[width=.8\linewidth]{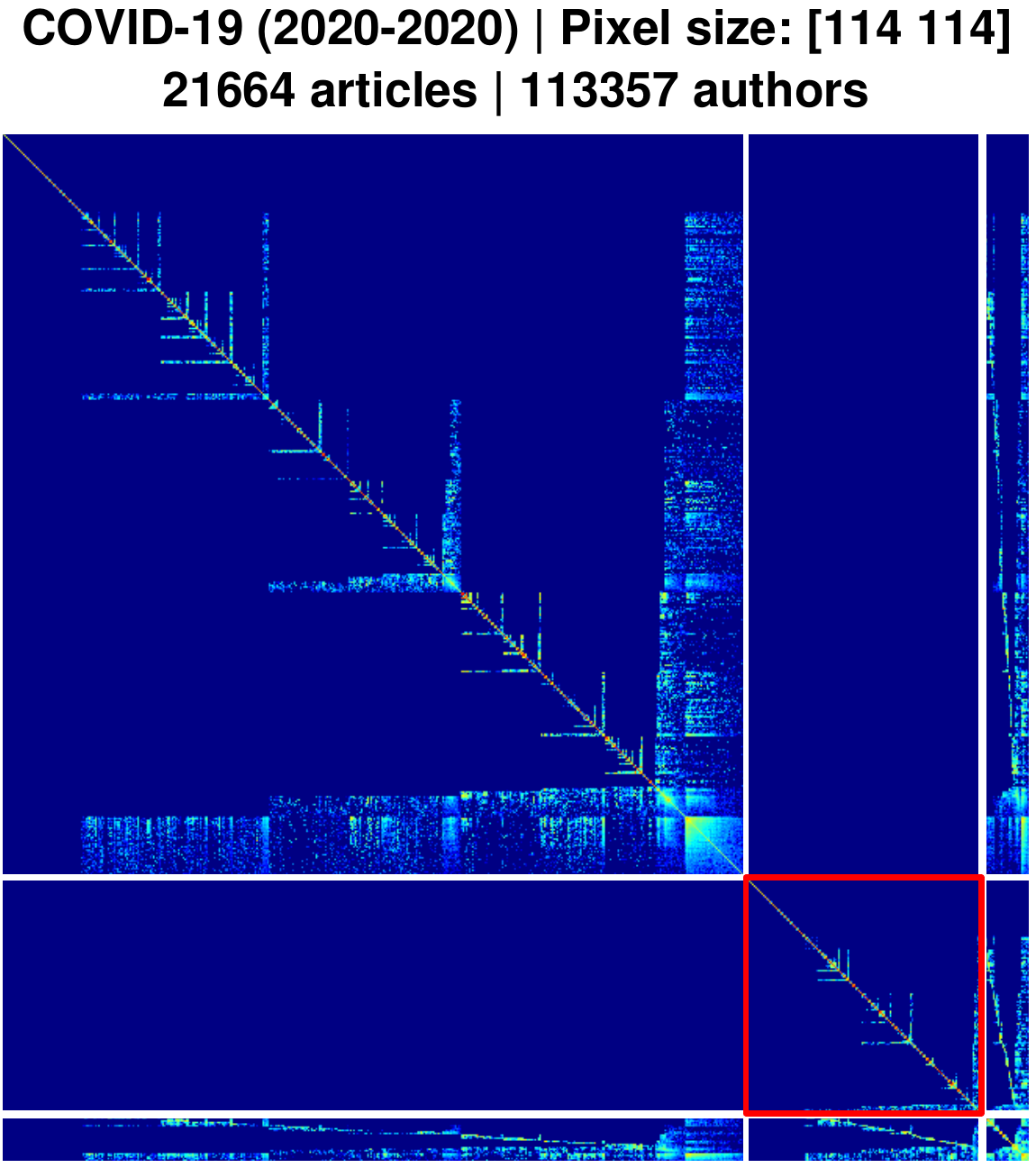}
  \end{subfigure}
  \begin{subfigure}{0.48\linewidth}
    \centering
    \includegraphics[width=.8\linewidth]{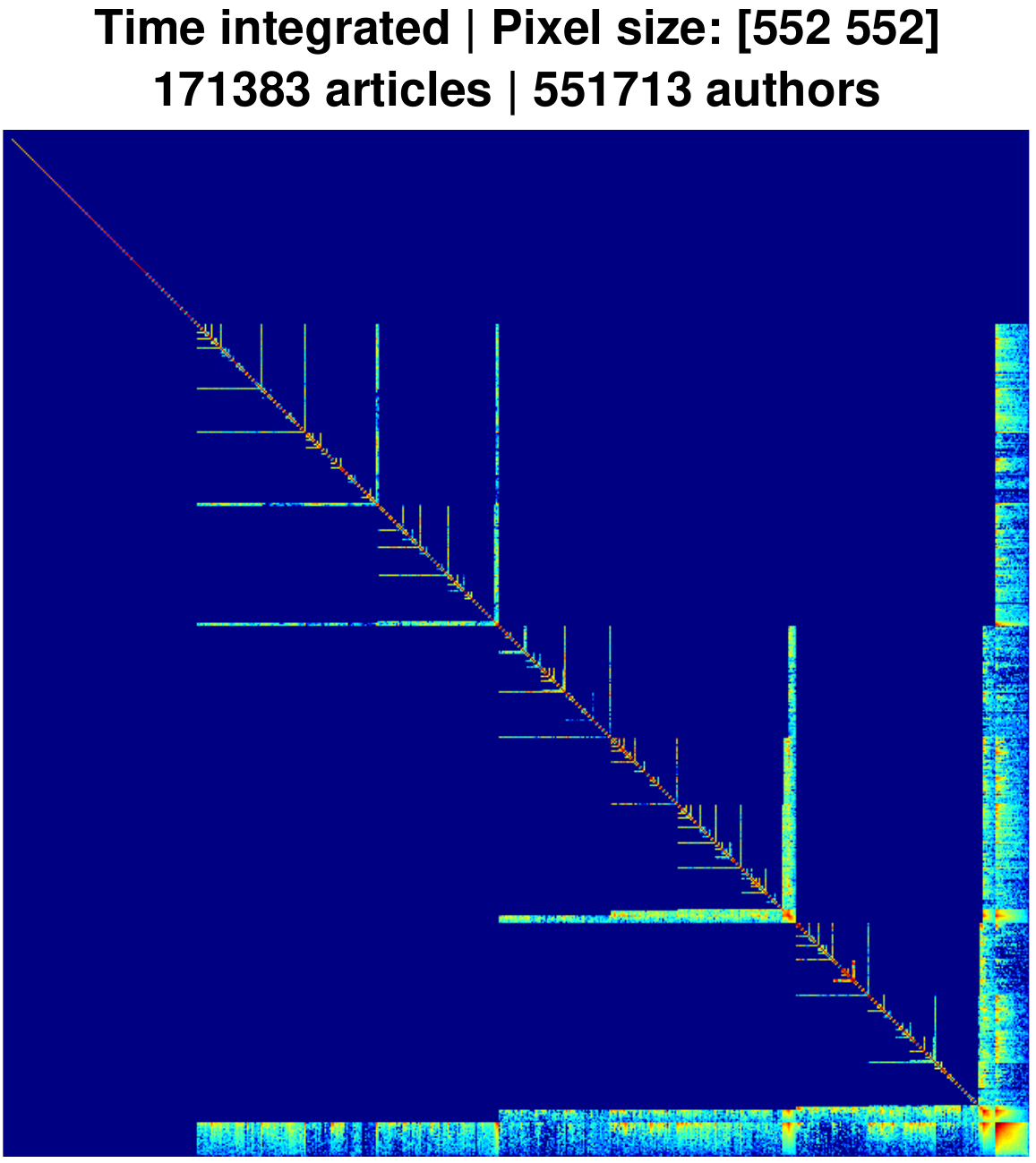}
  \end{subfigure}
  \caption{Four adjacency matrices for three epoch-associated
    collaboration networks and one time-integrated network up to the
    present, specified in the subtitles.
    In each matrix, the author vertices are first grouped by the
    septa-partition and ordering of (b) and (c) in \cref{fig:pipeline} as
    marked by {\bf white} partition lines; and then ordered within
    each cohort by combinatorial nested dissection (CND) in order to
    reveal the sparse topology.  The intra-collaboration in the {\tt
      Core} cohort (2) is marked by {\bf red} boundary lines.
    A  pixel in each display represents  a block matrix, as specified in the subtitle.
    The pixel color is determined by the number of links within the
    block, in logarithmic scale. The pixel in darker blue indicates a sparser block;
    in vibrant red, a denser block.
    \textbf{Observation}: The {\tt Cout} block (1) and {\tt Cin} block
    (3) represent precursors and followers, respectively.  The
    {Cout:Core} ratio is small in SARS, larger in MERS, and much
    larger in COVID-19.  Reversely, the {Cin:Core} ratio is the
    largest in SARS, showing the lasting impact. Such information is
    revealed via our approach with triad citation networks and
    septa-segmentation of collaboration networks.  }
  \label{fig:author-graph}
\end{figure}

\section{ Time-shifted \& convolved networks }
\label{sec:triad-subgraphs}

The conventional static collaboration network is constructed from the
article-author bipartite deprived of temporal activity information.
We introduce instead epoch-centered triad citation networks to enable
temporal pattern analysis at mixed time scales, in adaptation to
continuous and new collaboration activities.
We describe our model, its properties, computational procedure and
underlying rationale.

Let $T$ be a time window or epoch. Denote by {\tt Core} the set of
articles published in the period. See the citation matrix in (a) of
\cref{fig:pipeline}.  The core articles cite each other in the same
period, represented by the matrix block {\tt Core} on the
diagonal. This is a closed network. In reality, core articles cite
articles in set {\tt Cout} (outward links in the same column block),
and they are cited by articles in set {\tt Cin} (inward links in the
same row block). We term such open network as an {\em epoch-centered
  triad network}. We take the sub-bipartite with the article vertices
from the triad network on the one side and the involved authors on the
other. See the bipartite in (b) of \cref{fig:pipeline}.  We partition
the involved authors into seven cohorts. The \emph{septa-partition} is
shown in the Venn diagram in (c) of \Cref{fig:pipeline}. In
particular, the authors in cohort (4) are active not only during the
epoch $T$, but also before and after the epoch by one hop in
citation. We refer to this cohort as the {\em persistent cohort}.

Often, dynamic network analysis takes one of the extremes in dealing
with time scale\cite{newman2004,molontay2020,hofbauer1988}.  At the
one extreme, one assumes a closed network with fixed boundary and
a dynamics model that describes internal change at microscopic
time scale, subject to initial condition, boundary condition, and
some additional regulation condition. At the other extreme, a dynamic
network is sliced into multiple ones by non-overlapping time windows
at a macroscopic time scale. No time overlap nor memory/impact among
the sliced networks; no finer temporal resolution to differentiate
within each.  Each time-sliced network is then treated as static;
subsequent analysis across the networks is subject to the fixed time
resolution by the window size.

We reason differently.  Literature graphs are dynamic, but not on an
assumed uniform scale. Collaboration activities take place at
various and mixed time scales, similar to many social networks,
dissimilar to those physical-sensor networks with built-in clocks or
biological networks with intrinsic circadian rhythms.
Our model of epoch-centered triad citation networks is simple, and
innovative in using a data-adaptive meta resolution (e.g. adapt to
each epidemic period) in order to capture and accommodate the
variation in time resolution or scale.  By our schema, the networks
are open, not only time shifted but also permitting temporal
convolution and dilation.  The temporal convolution is by one-hop
topological links in citation to precursors and successors.  It is
time dilated in citation, not closed to or confined within an imposed
time window.  These properties of triad citation networks are
transported at ease to the collaboration networks, each of which is
consequently time-segmented within, at the meta time scale.
Based on the model, we are able to investigate connection,
continuation, differentiation, and deviation across networks centered
at different epochs.

\section{Differentiation of collaboration networks}
\label{sec:differentiation}

The conventional network characterization and correlation based on
degree distribution have limited capacity to reveal or differentiate
temporal and topological relations among time-segmented collaboration
networks, as shown evidently in \cref{fig:degree-histogram-sigma1}.
We introduce in this section how we differentiate the collaboration
networks at three granularity levels, and associate them as well, via
topological encoding with graphlets.

\subsection{Graphlet spectral descriptors}
\label{sec:graphlet-description}

\begin{figure}
  \centering
  \includegraphics[width=\linewidth]{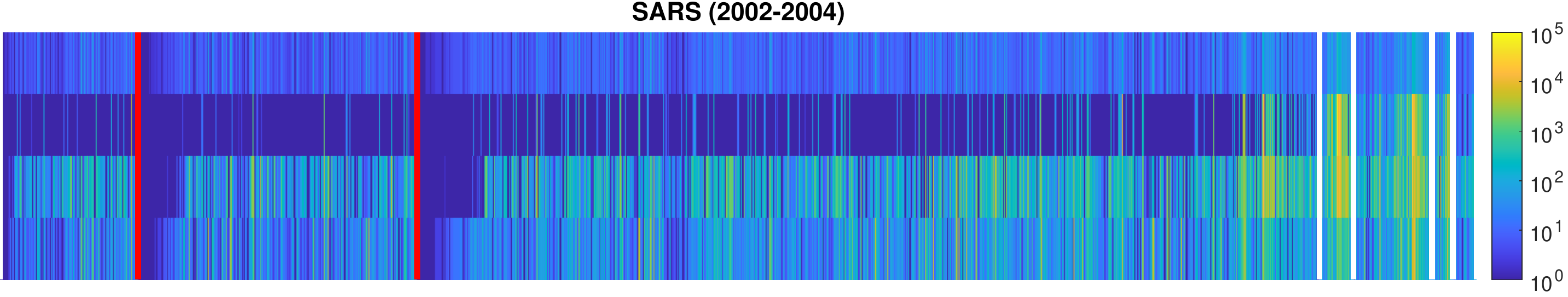}
  \\[0.5em]
  \includegraphics[width=\linewidth]{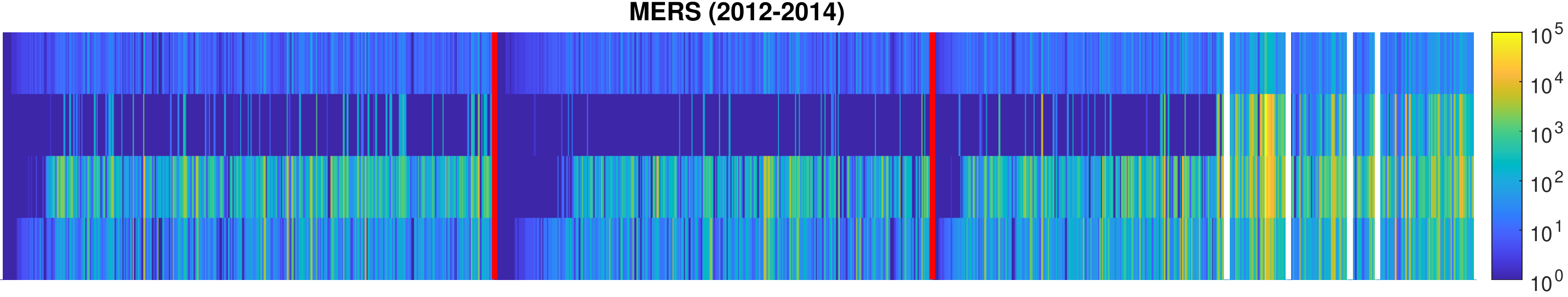}
  \\[0.5em]
  \includegraphics[width=\linewidth]{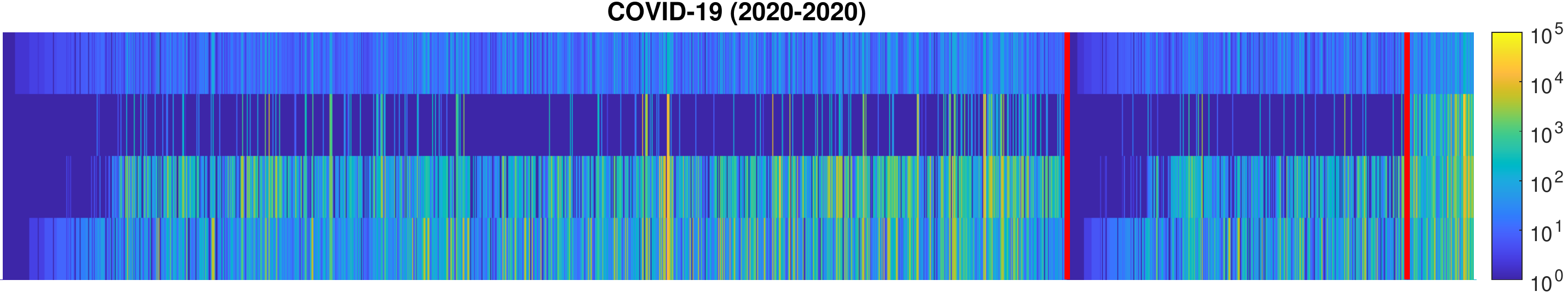}
  \\[0.5em]
  \caption{ %
    Graphlet spectrograms of three collaboration networks with
    respective labels and time windows as SARS (2002-2004), MERS
    (2012-2014), and COVID-19 (2020: Jan.-June).  See
    \Cref{eq:graphlet-transform} and the subsequent definition of
    graphlet spectrogram.  Each column identifies with an author
    vertex. Row $i$ contains $\sigma_i$-frequencies across the
    vertices, $i\! =\! 1\! :\! 4$. The frequencies with $\sigma_0$ are
    constantly $1$, not shown.  The ordering among the (author)
    vertices is the same as in the respective adjacency matrices in
    \cref{fig:author-graph}. The core set (2) is marked by red
    boundary lines. A pixel in brighter color indicates a higher
    frequency value.
    {\bf Observation}: In each of the top two spectrograms, the
    persistent cohort (4) has higher concentration of vertices with
    higher $\sigma_{2}$ frequencies and $\sigma_{3}$ frequencies. In
    fact, vertices with higher $\sigma_2$ frequencies have higher
    centrality positions, see \cref{sec:temporal-pattern-analysis}.  The
    persistent cohort (4) is missing in the spectrogram at the bottom,
    as the network is in its developing stage.}
  \label{fig:three-snapshot-spectrograms}
\end{figure}

We review generic graphlets and graphlet dictionaries by their forms
and attributes.  

A {\em graphlet} is a connected graph with a small vertex set and
a designated node to be incident with.  We show
in \Cref{fig:graphlets}  %
a dictionary of $5$ (undirected) graphlets,
$\Sigma_5 \! = \{\, \sigma_k \}_{k=0:4}$.
The dictionary contains small subgraph patterns: singleton, edge
($K_{2} $), binary fork ($K_{1,2}$), 2-path ($P_2$), and triangle
($C_3$, $K_3$).
The designated incidence node is shown with a red
square, up to an isomorphic permutation (shown in red circles).
Graphlets with the same number of nodes form a family with an
partial ordering.  For example, $\sigma_2$, $\sigma_3$ and $\sigma_4$
are a family of tri-node graphlets. The partial ordering
$\sigma_2, \sigma_3 \prec \sigma_4$ is by the relationship that
$\sigma_{2}$ and $\sigma_{3}$ are subgraphs of $\sigma_{4}$.

Specific to any undirected graph $G=(V,E)$, we obtain at every
vertex $v \in V$ a graphlet frequency vector of length $|\Sigma|$, the
element-$k$ of which is the number/frequency of $\sigma_k$-pattern
induced subgraphs that are incident to $v$,
$ k = 0, 1, \cdots, |\Sigma|\!-\!1$.
In other words, we make a transform of graph $G$ to a field of vectors
over $V$. The vectors encode, with graphlets as the coding
words/elements, the topological and statistical information of the
graph.
The dictionary $\Sigma_{2} = \{ \sigma_{0}, \sigma_{1} \}$ encodes the
very basic information. However, it limits network analysis to the
ordinary degree distributions, types, correlations and
models~\cite{newman2011,posfai2013}.
We use the dictionary $\Sigma_{5}$ with much greater coding capacity,
with little cost in computation. The relationship among the graphlets,
computation formulas, and complexities are detailed in
\cite{floros2020}.

We describe the graphlet frequencies in a graph $G=(V,E)$ with the
help of a vertex-graphlet incidence structure.
Denote by $ B = ( V, \Sigma ; E_{v\sigma}) $ the bipartite between the
graph vertices and the graphlets,
$E_{v\sigma} \subset V\times \Sigma$.
There is a link $(v,\sigma) $ between a vertex $v \in V$ and a
graphlet $\sigma \in \Sigma$ if $v$ is an incident node on an induced
subgraph of $\sigma$-pattern.
The incident node on a graphlet is uniquely specified, up to an
isomorphic mapping.
For example, graphlet $\sigma_4$ (clique $K_3$ or cycle $C_3$) in
\Cref{fig:graphlets} is an automorphism.
There may be multiple links between $v$ and $\sigma_k$.  We denote
them by a single link $(v, \sigma_k)$ with a positive integer weight
$d_k(v)$ for the multiplicity, which is the frequency with graphlet
$\sigma_k$.
However, the multiplicities from vertex $v$ to multiple graphlets in
the same family are not independently determined.
For example, the multiplicities on links from vertices to $\sigma_2$
do not include those to sub-graphlets within $\sigma_4$.  The weight on
$ (v, \sigma_1) $ is counted independently as $\sigma_1$ has no
other family member.
We describe formally the transformation of $G$ to the frequency vector
field over $V$, with $K = |\Sigma|-1$,
\begin{equation}
\label{eq:graphlet-transform}
\mathbf{f}( v ) =
\mathbf{f}( u | G) =  [ d_{0}(v), d_{1}(v), \cdots, d_{K}(v) ]^{\rm T}, 
\quad v \in V. 
\end{equation}
For any $k$, $ 0 \leq k < K $, $d_k(v)$ is the frequency of 
pattern-$\sigma_k$ subgraphs incident at $v$.
In particular, $d_{0}(v) = 1$, $d_1(v)$ is the degree of $v$ on graph
$G$. The descriptor $\mathbf{f}(v)$ encodes the topological
structure of the neighborhood of vertex $v$, with the graphlets
as the coding words/elements. 
We use the long-hand notation for node-wise descriptor
$\mathbf{f}(v | G)$ when necessary.
We present graphlet spectrograms of three epoch-associated
collaboration networks, labeled respectively as SARS, MERS and
COVID-19 in \Cref{fig:three-snapshot-spectrograms}. The 
descriptors are color-coded and placed in columns, with nodes in the
same ordering as in \Cref{fig:author-graph}.
The graphlet dependencies, fast and exact transform formulas and
complexity analysis are detailed in~\cite{floros2020b}.
In implementation, we use {\tt GraphBLAS}~\cite{davis2018} 
to exploit sparse patterns and Cilk~\cite{blumofe1996} for multi-threaded
programming.

\begin{table}
  \centering
  \caption{Pairwise agreement scores by $1\!-\! \eta$, in the upper
    triangular, among $6$ epoch-centered collaboration networks,
    $\eta$ is defined in \cref{eq:discrepancy-measure-XY}.  In the
    lower triangular are the sizes of pairwise intersections among the
    vertex sets, by the unit of a thousand. Observations are
    elaborated in \cref{sec:temporal-pattern-analysis}.  }
\resizebox{\linewidth}{!}{%
\begin{tabular}{|l|r|r|r|r|r|r|}
\hline
  & \multicolumn{1}{c|}{SARS}        & \multicolumn{1}{c|}{Swine flu}
  & \multicolumn{1}{c|}{MERS}       & \multicolumn{1}{c|}{Ebola}
  & \multicolumn{1}{c|}{Avian flu} & \multicolumn{1}{c|}{COVID-19} \\
\hline
SARS & - & \textbf{0.68} & \textbf{0.65} & \textbf{0.62} & \textbf{0.54} & \textbf{0.33} \\
\hline
Swine flu & \SI{145}{\kilo\nothing} & - & \textbf{0.75} & \textbf{0.71} & \textbf{0.61} & \textbf{0.35} \\
\hline
MERS & \SI{143}{\kilo\nothing} & \SI{206}{\kilo\nothing} & - & \textbf{0.79} & \textbf{0.63} & \textbf{0.39} \\
\hline
Ebola & \SI{138}{\kilo\nothing} & \SI{199}{\kilo\nothing} & \SI{236}{\kilo\nothing} & - & \textbf{0.65} & \textbf{0.42} \\
\hline
Avian flu & \SI{120}{\kilo\nothing} & \SI{171}{\kilo\nothing} & \SI{188}{\kilo\nothing} & \SI{199}{\kilo\nothing} & - & \textbf{0.49} \\
\hline
COVID-19 & \SI{35}{\kilo\nothing} & \SI{46}{\kilo\nothing} & \SI{53}{\kilo\nothing} & \SI{60}{\kilo\nothing} & \SI{73}{\kilo\nothing} & - \\
\hline
\end{tabular}%
}
\vspace*{-15pt}
\label{tab:eta_intersection-size}
\end{table}

\subsection{Discrepancy measures}
\label{sec:discrepancy-measure}

\begin{table*}
  \centering
  \caption{Agreement scores by $1\!-\!\eta$, at the cohort level, between
    MERS as the comparison target and the other networks as the references, along
    with aggregation weights (in red) of
    \Cref{eq:discrepancy-measure-on-X}.  %
    In the left table, ${\cal X} = \{X_i\}_{i=1:7}$ is the
    septa-cohort-partition for the target MERS, the scores are
    $1\!-\! \eta(X_i, G_{z})$, where $G_{z}$ changes across the
    columns associated respectively with the reference networks. 
    In the right table, ${\cal Y}$ is the septa-cohort-partition of
    each target network, the scores (in black) are
    $1\!-\! \eta(Y_j, G_z)$.  The weights (in red) are based on the
    cohort sizes in \Cref{tab:cohort-sizes}.  The last rows in both
    tables are the components of \cref{eq:discrepancy-measure-XY}.
    The sums give entries of
    \cref{tab:eta_intersection-size}, which summarizes all such comparisons at cohort level.
    {\bf Key observation.} The cohorts of MERS mostly remain in
    network Ebola (by the left table), and are more influential to the developing network
    COVID-19 (by the right table).  }
  \begin{subfigure}{0.20 \linewidth}
    \centering
    \includegraphics[height=0.11\textheight]{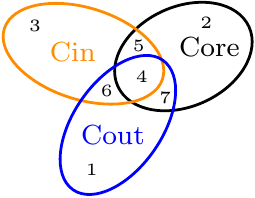}
  \end{subfigure}
  \hskip 1em
  \resizebox{0.34 \linewidth}{!}{%
  \begin{tabular}{|c|r|r|r|r|r|r|}
    \hline
     & $w_i^x$ & \multicolumn{1}{c|}{SARS} &\multicolumn{1}{c|}{Swine flu} & \multicolumn{1}{c|}{Ebola} & \multicolumn{1}{c|}{Avian flu} & \multicolumn{1}{c|}{COVID-19} \\
    \hline
    1 & \textcolor{red}{0.34} & 0.59 & 0.83 & 0.78 & 0.58 & 0.24 \\
    \hline
    2 & \textcolor{red}{0.30} & 0.39 & 0.51 & 0.73 & 0.53 & 0.32 \\
    \hline
    3 & \textcolor{red}{0.20} & 0.52 & 0.75 & 0.92 & 0.87 & 0.36 \\
    \hline
    4 & \textcolor{red}{0.04} & 0.60 & 0.85 & 0.93 & 0.82 & 0.21 \\
    \hline
    5 & \textcolor{red}{0.04} & 0.44 & 0.70 & 0.91 & 0.79 & 0.19 \\
    \hline
    6 & \textcolor{red}{0.02} & 0.55 & 0.85 & 0.89 & 0.77 & 0.12 \\
    \hline
    7 & \textcolor{red}{0.07} & 0.49 & 0.78 & 0.82 & 0.61 & 0.12 \\
    \hline
    $1\! -\! \eta_{\cal X}$ & - & 0.50 & 0.71 & 0.81 & 0.65 & 0.28 \\
    \hline
  \end{tabular}%
}
\hskip 0.5em
\resizebox{0.41 \linewidth}{!}{%
\begin{tabular}{|c|r|r|r|r|r|}
\hline
 & \multicolumn{1}{c|}{SARS} & \multicolumn{1}{c|}{Swine flu} & \multicolumn{1}{c|}{Ebola} & \multicolumn{1}{c|}{Avian flu} & \multicolumn{1}{c|}{COVID-19} \\
\hline
1 & \textcolor{red}{0.09} \quad 0.71 & \textcolor{red}{0.24} \quad 0.73 & \textcolor{red}{0.39} \quad 0.87 & \textcolor{red}{0.46} \quad 0.79 & \textcolor{red}{0.72} \quad 0.51 \\
\hline
2 & \textcolor{red}{0.19} \quad 0.58 & \textcolor{red}{0.25} \quad 0.63 & \textcolor{red}{0.35} \quad 0.63 & \textcolor{red}{0.40} \quad 0.40 & \textcolor{red}{0.23} \quad 0.46 \\
\hline
3 & \textcolor{red}{0.61} \quad 0.87 & \textcolor{red}{0.34} \quad 0.90 & \textcolor{red}{0.11} \quad 0.83 & \textcolor{red}{0.03} \quad 0.71 & \phantom{0.00} \quad \phantom{-...-}- \\
\hline
4 & \textcolor{red}{0.02} \quad 0.86 & \textcolor{red}{0.04} \quad 0.91 & \textcolor{red}{0.03} \quad 0.89 & \textcolor{red}{0.00} \quad 0.74 & \phantom{0.00} \quad \phantom{-...-}- \\
\hline
5 & \textcolor{red}{0.05} \quad 0.81 & \textcolor{red}{0.06} \quad 0.88 & \textcolor{red}{0.03} \quad 0.75 & \textcolor{red}{0.01} \quad 0.47 & \phantom{0.00} \quad \phantom{-...-}- \\
\hline
6 & \textcolor{red}{0.01} \quad 0.84 & \textcolor{red}{0.02} \quad 0.88 & \textcolor{red}{0.01} \quad 0.91 & \textcolor{red}{0.00} \quad 0.79 & \phantom{0.00} \quad \phantom{-...-}- \\
\hline
7 & \textcolor{red}{0.02} \quad 0.69 & \textcolor{red}{0.04} \quad 0.76 & \textcolor{red}{0.08} \quad 0.84 & \textcolor{red}{0.10} \quad 0.68 & \textcolor{red}{0.05} \quad 0.46 \\
\hline
$1\! -\! \eta_{\cal Y}$ & \phantom{x.xxx} \quad 0.79 & \phantom{x.xxx} \quad 0.79 & \phantom{x.xxx} \quad 0.78 & \phantom{x.xxx} \quad 0.62 & \phantom{x.xxx} \quad 0.49 \\
\hline
\end{tabular}%
}
\vspace*{-10pt}
\label{tab:discrepancy-planB-MERS}
\end{table*}

\begin{table}
  \centering
  \caption{Sizes of $7$ author cohorts in each of the $6$
    collaboration networks associated with the labeled epochs
    of \Cref{tab:epochs},  with cohorts grouped by the
    septa-partition of (c) in \Cref{fig:pipeline}. 
    They are used in the weighting scheme for topological discrepancy
    aggregation of \cref{eq:discrepancy-measure-on-X} and
    \cref{tab:eta_intersection-size,tab:discrepancy-planB-MERS}.
    The size of the largest connected component is in the last row.
    {\bf Observation.} The total number of authors with each of the
    networks increases steadily with the forward shift in the
    epochs. We expect a burst with COVID-19 as the network
    already has a massive size while in its early developing stage
    (and therefore the absence of cohorts (3)-(6)).  }
  \resizebox{0.85 \linewidth}{!}{%
    \begin{tabular}{|l|r|r|r|r|r|r|}
      \hline
      & \multicolumn{1}{c|}{SARS} & \multicolumn{1}{c|}{Swine flu}
      & \multicolumn{1}{c|}{MERS} & \multicolumn{1}{c|}{Ebola}
      & \multicolumn{1}{c|}{Avian flu} & \multicolumn{1}{c|}{COVID-19} \\
      \hline
      1 & \num{16997} & \num{65214} & \num{98165} & \num{118386} & \num{143909} & \num{81939} \\
      \hline
      2 & \num{34354} & \num{66524} & \num{86934} & \num{105768} & \num{124864} & \num{26074} \\
      \hline
      3 & \num{111047} & \num{90654} & \num{58352} & \num{33863} & \num{7934} & \num{0} \\
      \hline
      4 & \num{4164} & \num{11648} & \num{12243} & \num{9663} & \num{1462} & \num{0} \\
      \hline
      5 & \num{9661} & \num{14847} & \num{12205} & \num{9094} & \num{1879} & \num{0} \\
      \hline
      6 & \num{2571} & \num{6204} & \num{5494} & \num{3589} & \num{481} & \num{0} \\
      \hline
      7 & \num{2922} & \num{11727} & \num{19628} & \num{25479} & \num{32727} & \num{5344} \\
      \hline
      $|V|$ & \num{181716} & \num{266818} & \num{293021} & \num{305842} & \num{313256} & \num{113357} \\
      \hline
      LCC & \num{143653} & \num{219982} & \num{240560} & \num{248824} & \num{249110} & \num{78006} \\
      \hline
    \end{tabular}%
  }
  \vspace*{-15pt}
  \label{tab:cohort-sizes}
\end{table}

Let $G_{x} = ( V_{x} , E_{x} )$ and $ G_{y} = ( V_{y} , E_{y} ) $ be
two collaboration networks.
By the septa-partition (c) of \Cref{fig:pipeline}, the (author)
vertices on each collaboration network is partitioned into $7$
non-overlapping cohort clusters. Denote by
${\cal X} = \{ X_i, i=1\! : \! 7 \}$ the septa-partition of $V_x$; by
${\cal Y} = \{ Y_j, j=1:7 \} $, of $V_y$.
Denote by $G_{z} $ their intersection, i.e.,
$ G_{z} = G_{x} \cap G_{y} = ( V_{x} \cap V_{y} , E_{x} \cap E_{y} )
$. The graph intersection is the largest subgraph shared by $G_x$ and
$G_y$.
The symmetric difference between the graph vertex sets is
$ V_{x} \Delta V_{y}= (V_{x} - V_{y}) \cup (V_{y} - V_{x}) = ( V_{x} -
V_{z} ) \cup (V_y - V_{z} ) $.
If $ V_{x} \Delta V_{y} \neq \emptyset$, there are authors active in
one network, inactive in the other.
Yet otherwise, i.e., $V_{x} = V_{y}$, the two networks $G_{x}$ and $G_
{y}$ may still differ in topological connections, rendering non-empty
symmetric difference between the edge sets, $ E_x
\Delta E_{y} = (E_x - E_{z}) \cup ( E_{y} - E_{z})$.
However, the size of the symmetric difference between the edge sets
lacks the capacity to characterize and differentiate the changes in collaboration patterns. 

We measure the discrepancy (discordance), or agreement (concordance),
in topological structures between $G_{x}$ and $G_{y}$, using the
graphlet spectral descriptors.
We measure the discrepancy at three granularity levels.
Define the relative difference between two vectors $\mathbf{a}$ and
$\mathbf{b}$ as
$ \mbox{rdiff} = | \mathbf{a} - \mathbf{b} | ./ | \mathbf{a} +
\mathbf{b} | $, where $./$ denotes the Hadamard division, i.e.,
element-wise division.
Let $G'$ be a graph. If $ v \notin V(G') $, we set
$ \mathbf{f}( v | G' ) = 0 $.
We measure the relative difference in topological connectivity at each
vertex $v$ between $G_{x}$ and $G_{z}$,
\begin{equation}
\label{eq:node-neighborhood}
  \eta( v , V_{x}, G_{z} ) \!= \big| \, \mbox{rdiff} 
  \big( \, \mathbf{f}( v | G_{x}), \mathbf{f}( v | G_{z} ) \, \big ) \, \big|_{p} ,
  \quad v \in V_{x} , 
\end{equation}
where $| \mathbf{a} |_{p}$ is the order-$p$ H{\"o}lder mean of vector
$\mathbf{a}$ with its elements in absolute values. The weights are
predetermined on the graphlets in the dictionary; equal weights are
used for empirical study later.
Strictly speaking, the difference
$\mathbf{f} ( v | G_{x} ) - \mathbf{f} ( v | G_{z} )$ characterizes
the change in the {\em topological neighborhoods} of vertex $v$ in
graph $G_x$ and graph $G_{z}$.
The first element, associated with the singleton graphlet
$\sigma_{0}$, accumulates to the difference in the vertex set between
$V_x$ and $V_{z}$. The second element reflects the difference in the
ordinary degree at every vertex.
When graphlet $\sigma_k$, $k>0$, is absent in both vertex
neighborhoods, element-$k$ in the relative difference is zero.
We aggregate the local changes to the vertex subsets by the
 septa-partition and then to the entire vertex set of $G_{x}$,
\begin{subequations}
  \label{eq:discrepancy-measure-on-X}
  \begin{align}
    \label{eq:discrepancy-measure-on-X-a}
    \eta( X_{i} , G_{z}  )=
    & 
      \frac{1}{|X_i|}    \sum_{v \in X_{i} } \eta( v, V_{x}, G_{z} ) , 
    \quad 1 \leq i \leq 7 , 
    \\
    \label{eq:discrepancy-measure-on-X-b}
    \eta( {\cal X} ,  G_{z}  )=
    & 
    \sum_{i=1:7}  w_i^{x}  \cdot \eta(  X_i,  G_{z} )  .
   \end{align} 
 \end{subequations}
 The weights in \Cref{eq:discrepancy-measure-on-X-a} are equal.  The
 weights in \Cref{eq:discrepancy-measure-on-X-b} increase
 monotonically with $ |X_i|/|V_{x}| $ and sum to $1$.

\begin{figure}
  \centering
  \includegraphics[width=.32\linewidth]{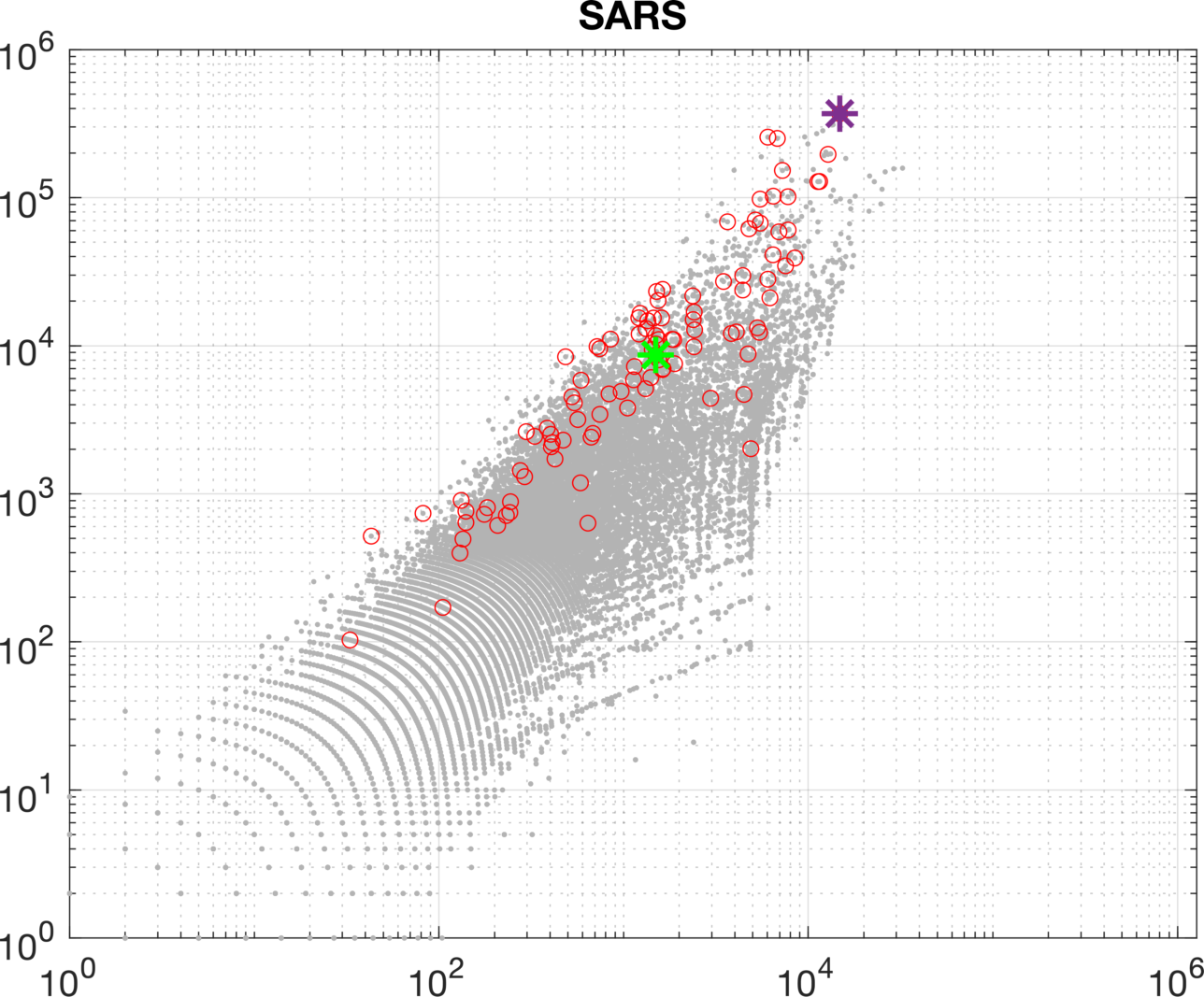}
  \includegraphics[width=.32\linewidth]{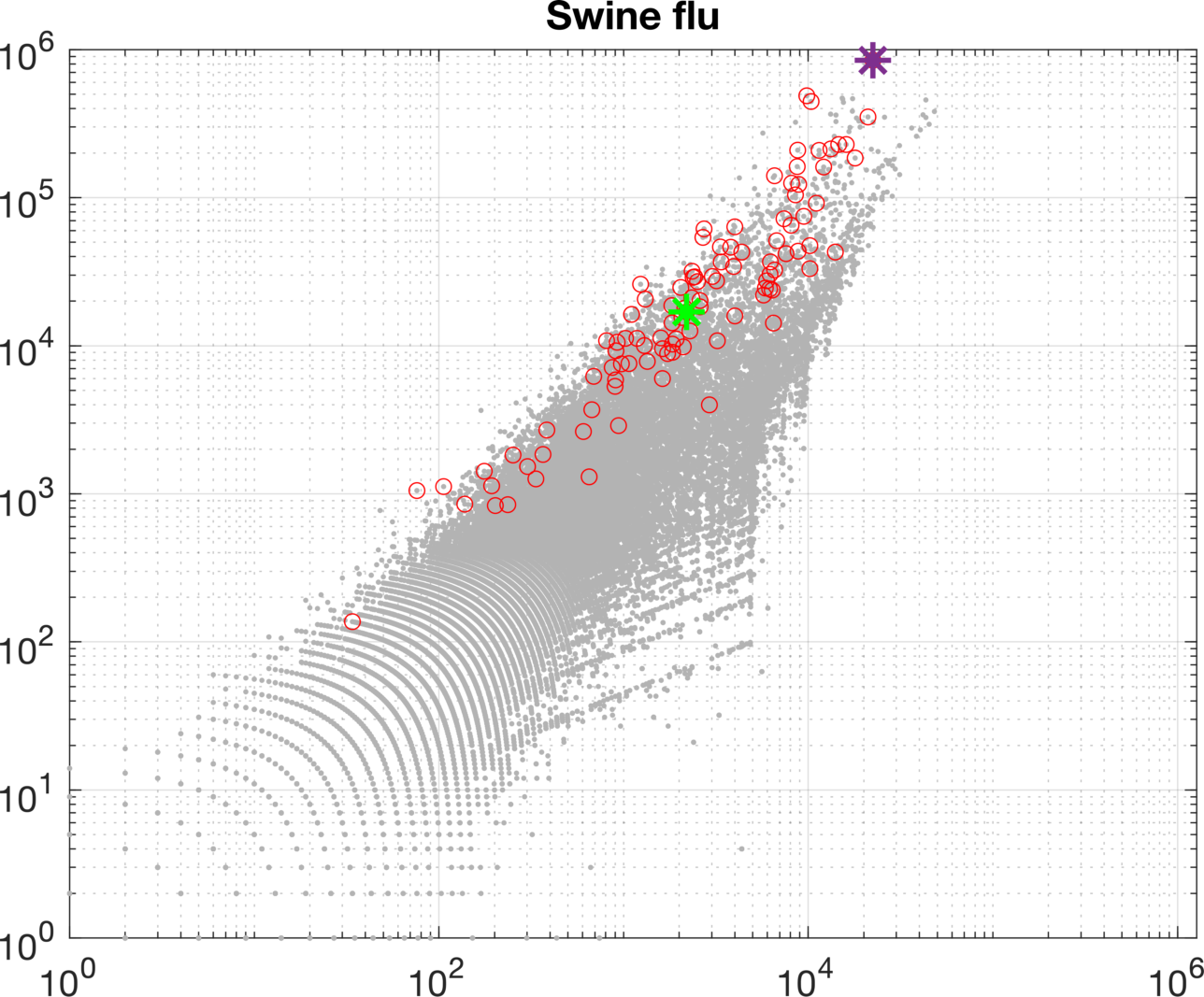}
  \includegraphics[width=.32\linewidth]{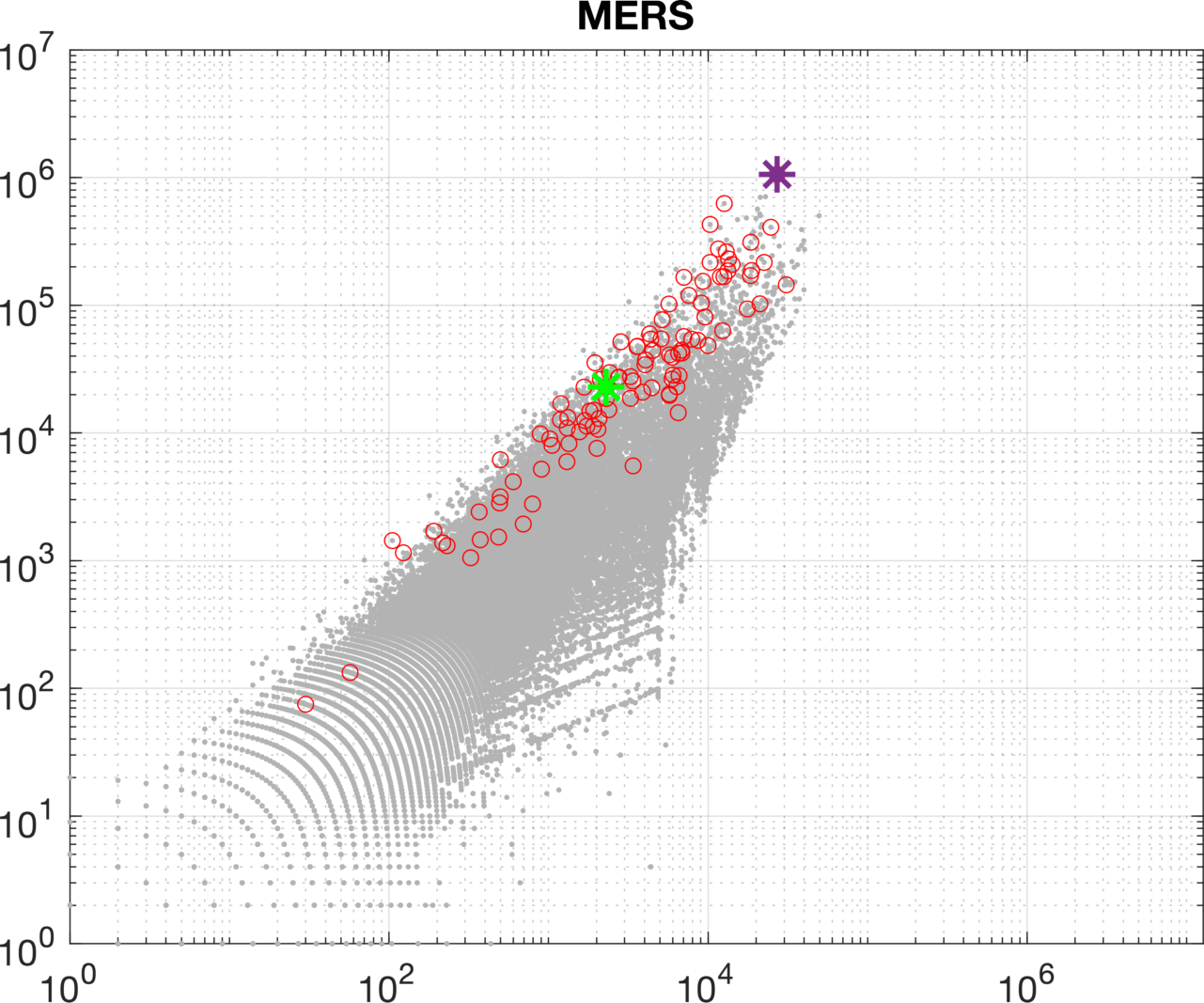}
  \\[0.5em]
  \includegraphics[width=.32\linewidth]{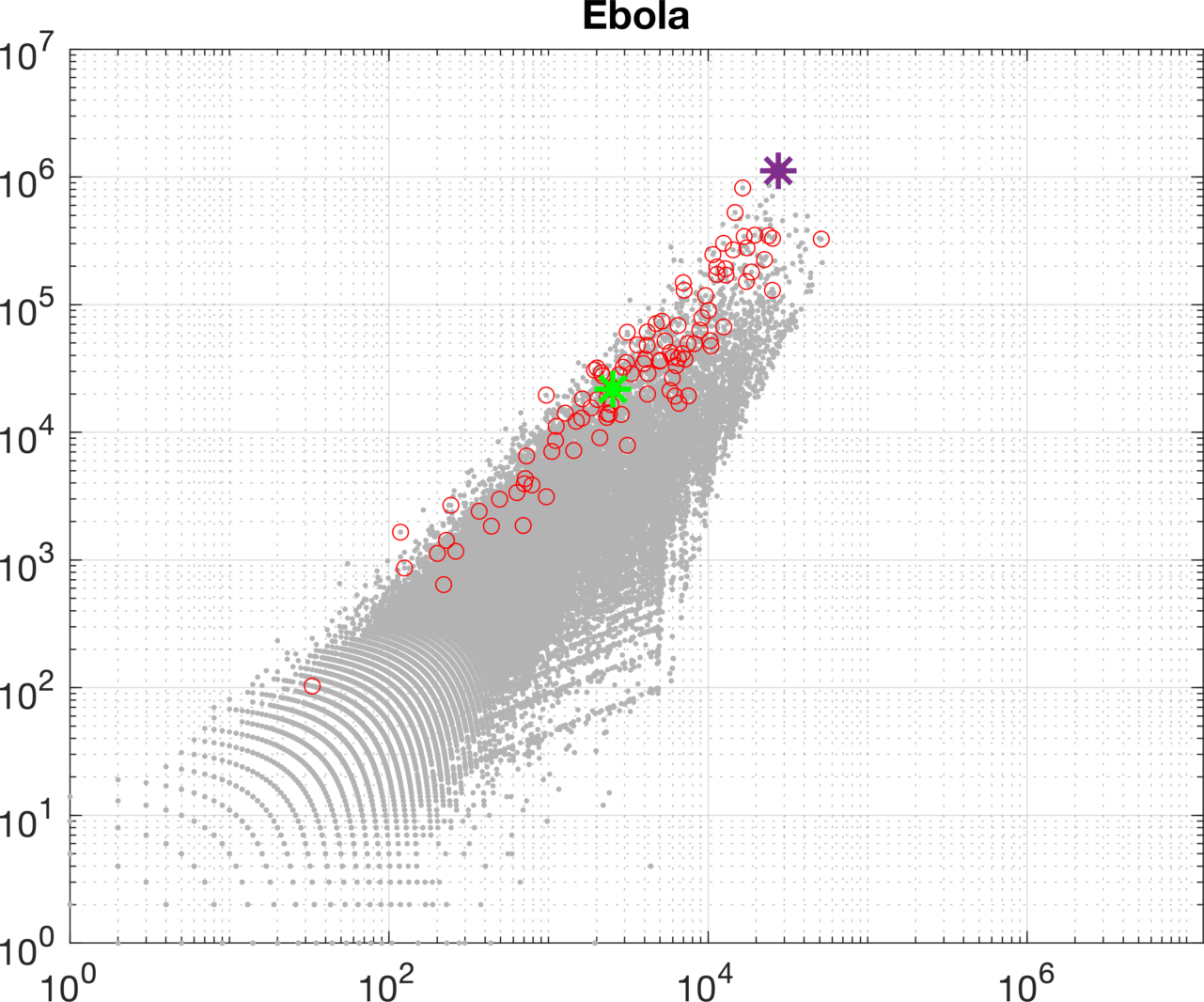}
  \includegraphics[width=.32\linewidth]{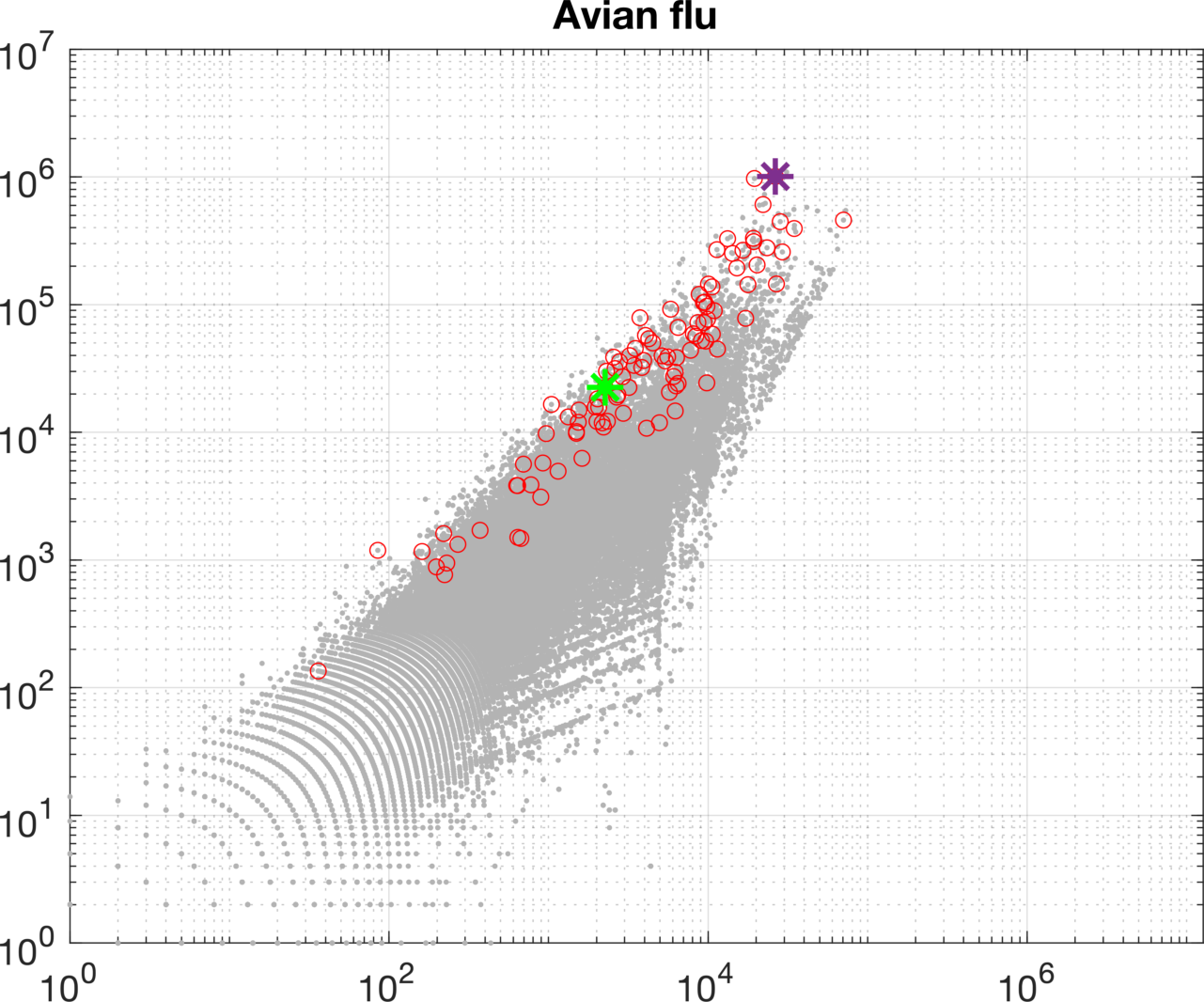}
  \includegraphics[width=.32\linewidth]{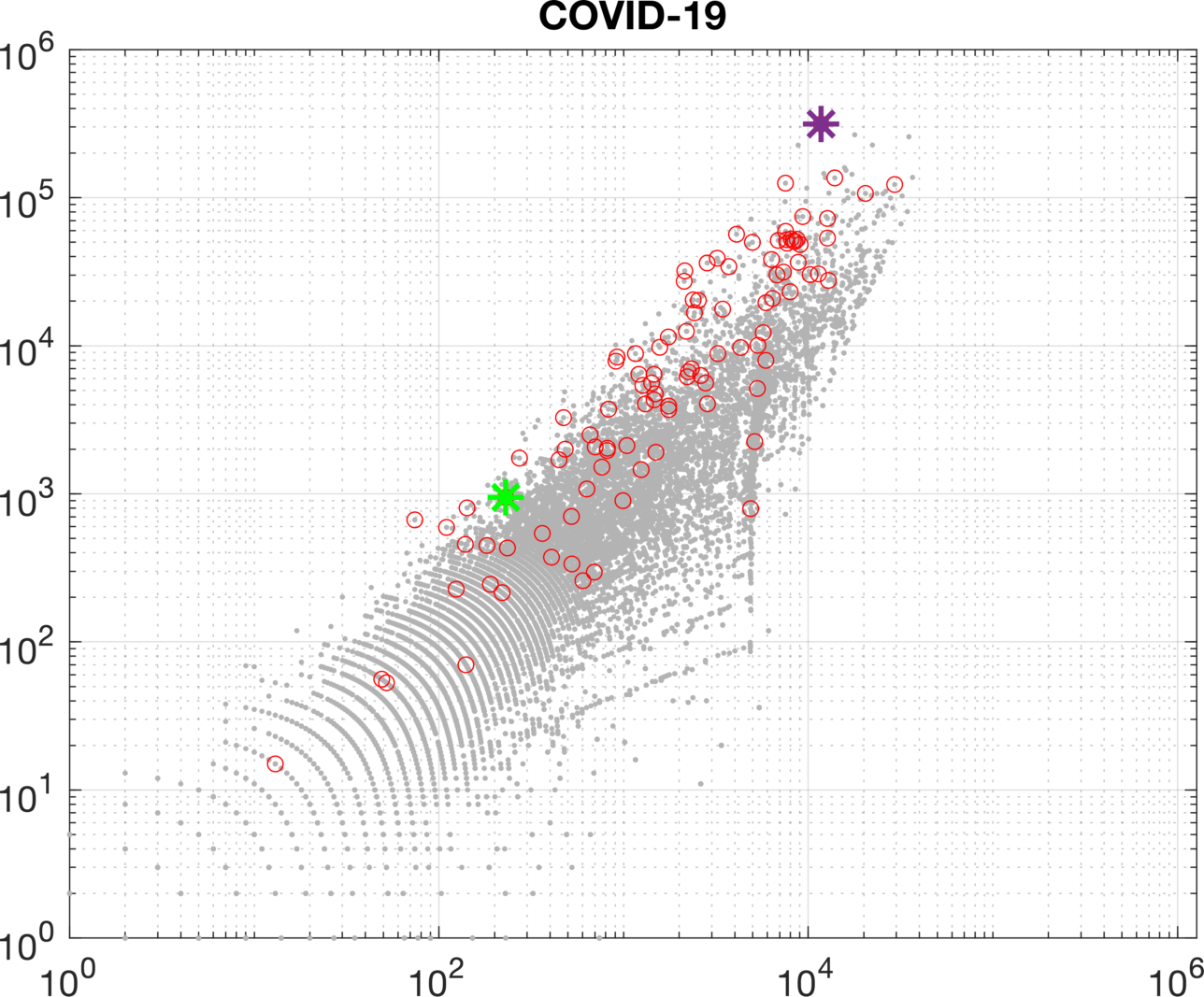}
  \\[0.5em]
  \caption[Scatter plots]{%
    Graphlet frequencies of $\sigma_2$ (bi-fork $K_{1,2}$) against
    $\sigma_{4}$ (cycle $C_3$), in logarithmic scatter plots, among
    the author vertices in the persistent cohort
    (4), one for each of the $6$ collaboration networks, except that
    \texttt{Cin} is empty in network COVID-19.
    See \cref{tab:cohort-sizes} for their respective sizes.  %
    The points in red are common to all $6$ persistent cohorts across
    the networks, they distinguish and highlight \textcolor{red}{103} 
    researchers who are continuously active and influential in virus
    research through two decades or longer.  Among them are
    {S. J. M. Peiris} (at the top in darker red) and {A. S. Fauci} (in
    bright green).
    {\bf Observation.}
    First, the dynamic persistence information lost in a static collaboration
    network and analysis  is uncovered by our novel analysis method.
    Secondly, we note in particular the phenomenon, common to
    all 6 networks, that the authors in red tend to have relatively
    higher $\sigma_2$-frequencies than the others in gray in the
    same local range of $\sigma_4$-frequencies.
    See further elaboration in \cref{sec:temporal-pattern-analysis}.
  } %
  \label{fig:scatter-plots-sigma42}
\end{figure}

 Similarly, we define $ \eta( v, V_{y}, G_{z} ) $,
 $\eta( Y_j, G_{z} ) $, $ 1 \leq j \leq 7 $, and
 $ \eta( {\cal Y} , G_{z} ) $.  We note the inequalities at many
 vertices, $ \eta( v, V_x, G_{z} ) \neq \eta ( v, V_y, G_{z} ) $,
 unless the change in topology between $G_{x}$ and $G_{y}$ is small.
 The aggregation weights are symmetrical between ${\cal X}$ and
 ${\cal Y}$ in the sense that $\{ w_{i}^{x} \}$ are the same as
 $\{ w_{i}^{y} \}$ when $|X_i| = |Y_{i}|$, $ 1 \leq i \leq 7 $.
 Together, we have a symmetric scalar measure of the topological
 discrepancy between $G_{x}$ and $G_{y}$,
\begin{equation}
  \label{eq:discrepancy-measure-XY}
  \eta( G_{x},  G_{y}  ) = \eta ( {\cal X},  G_{z}) /2 + \eta( {\cal Y}, G_{z} )  /2 . 
\end{equation}
The discrepancy value $\eta(  G_{x}, G_{y} ) $ is zero if and only
if $G_{x} = G_{y}$. It is positive and bounded by $1$ otherwise.

\section{Findings from uncovered temporal patterns} 
\label{sec:temporal-pattern-analysis}

\begin{figure}[!t]
  \centering
  \includegraphics[width=\linewidth]{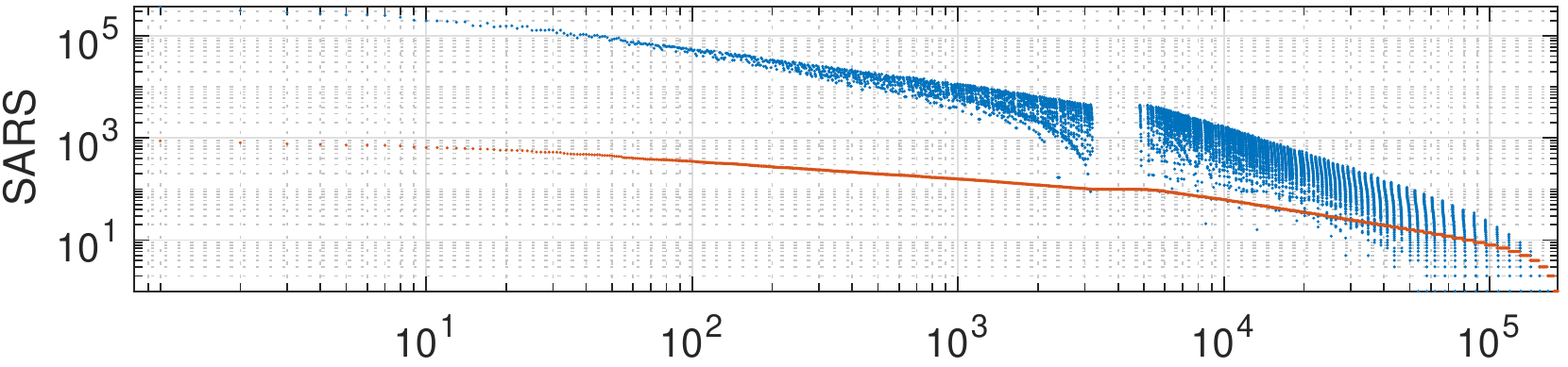}
  \\[0.5em]
  \includegraphics[width=\linewidth]{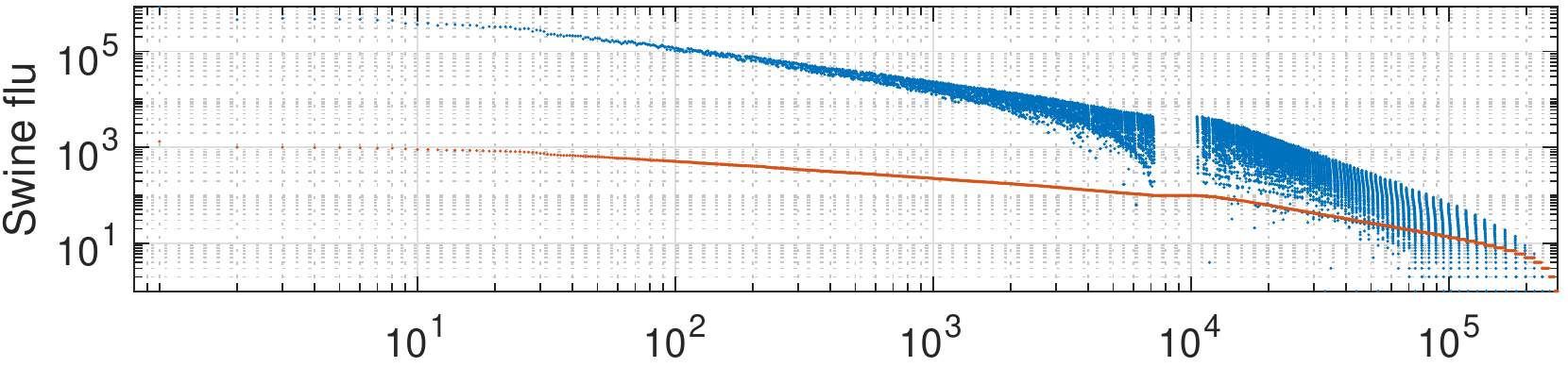}
  \\[0.5em]
  \includegraphics[width=\linewidth]{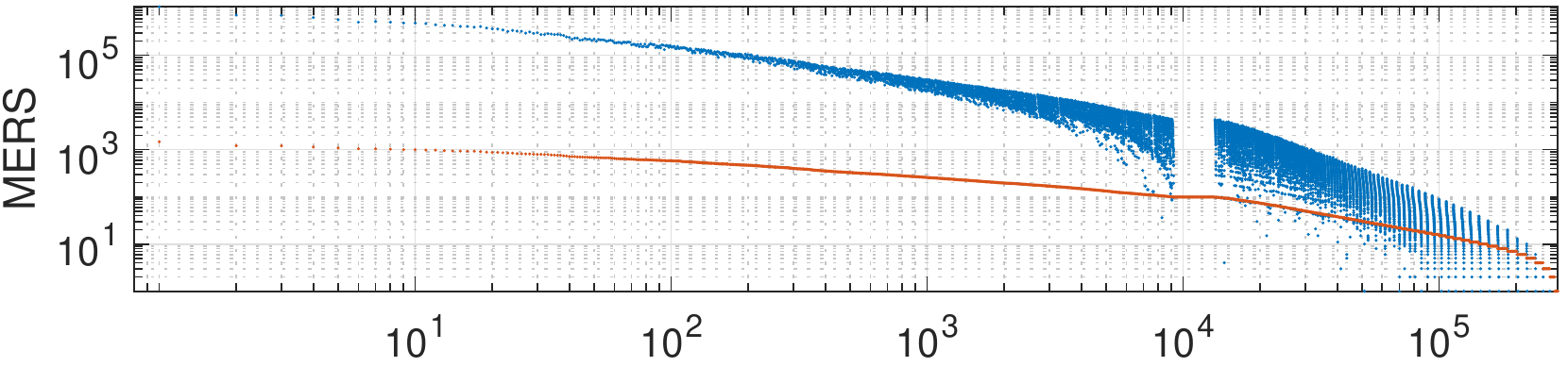}
  \\[0.5em]
  \includegraphics[width=\linewidth]{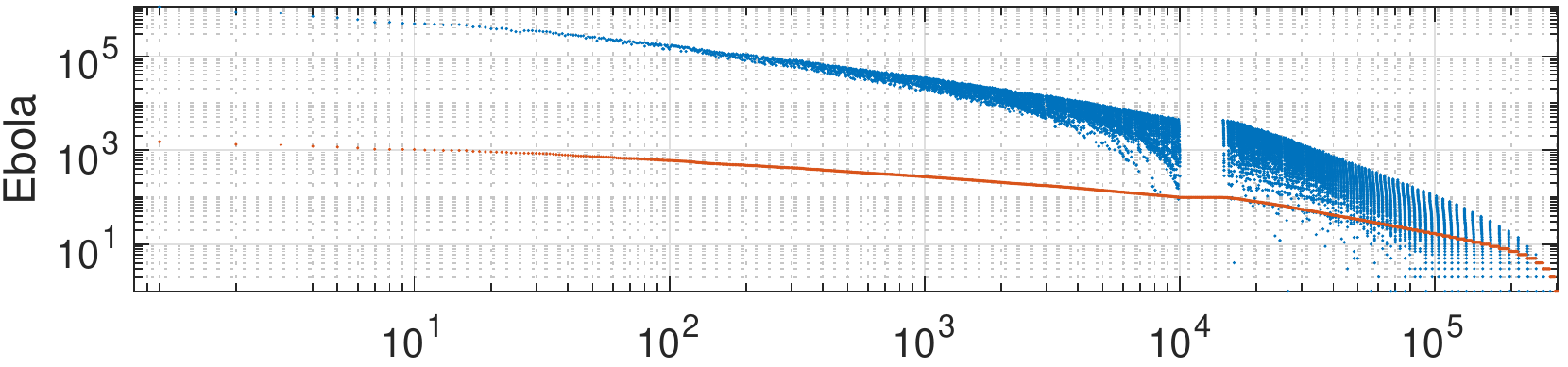}
  \\[0.5em]
  \includegraphics[width=\linewidth]{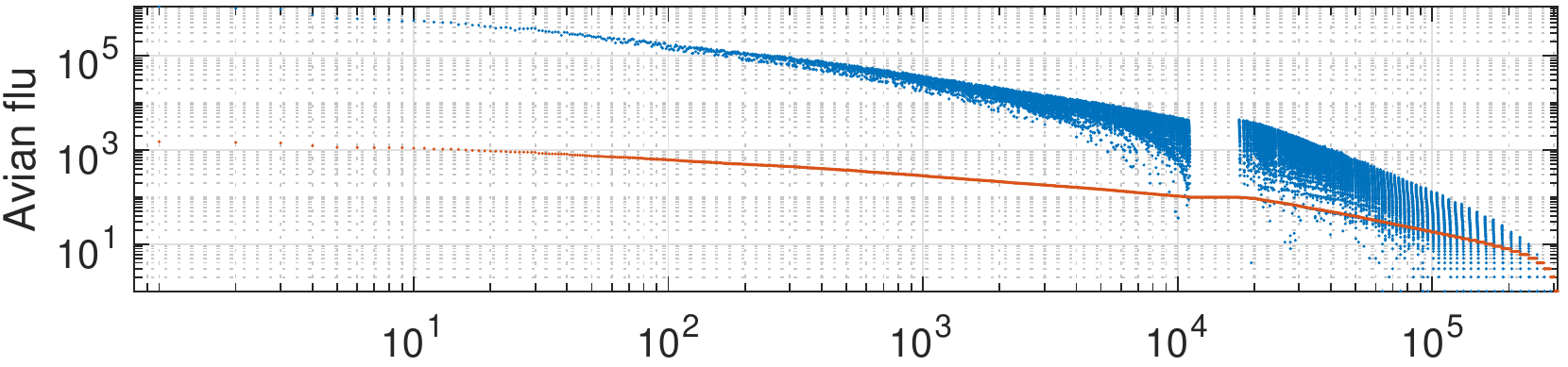}
  \\[0.5em]
  \includegraphics[width=\linewidth]{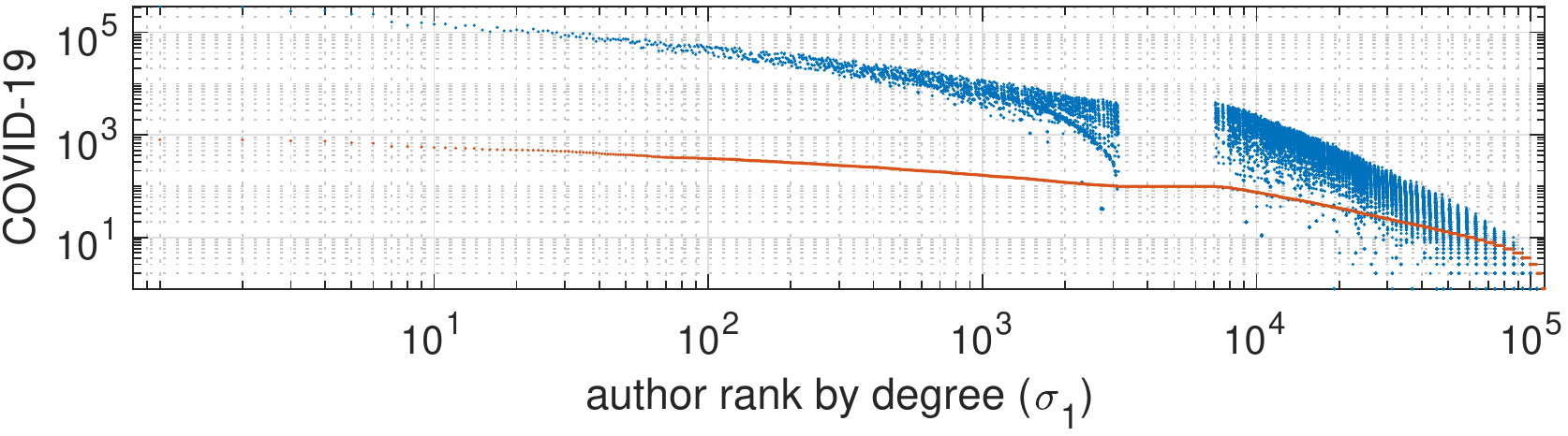}
  \\[-0.5em]
  \caption{%
    Topological betweenness among multiple collaboration clusters
    captured and encoded by $\sigma_2$-frequencies, which are shown in
    blue points placed by the rank-size distribution of degrees
    ($\sigma_1$) in red.  %
    See \Cref{sec:temporal-pattern-analysis} about the betweenness of
    $\sigma_2$, the bi-folk graphlet ($K_{1,2}$).  %
    The plot for each of the $6$ epoch-centered collaboration networks
    is of log-log scales.  %
    {\bf Key observation.}  %
    (i) Author nodes with the same degree of $\sigma_1$ (in $y$
    coordinate) differ greatly in $\sigma_2$-frequencies by over an
    order of magnitude.  An author with higher $\sigma_2$-frequency
    makes more connections among different author (triangle) clusters on
    different articles.
    (ii) Nodes with higher $\sigma_{2}$-frequencies include the hub-nodes
    (with top degrees)  but also significantly outnumber in accumulated
    $\sigma_2$-frequencies those of the hub-nodes by orders of
    magnitude.  They are responsible for the grass-root connectivity
    of a network.   %
    (iii) The difference between $\sigma_2$ and $\sigma_1$ is 
    incidentally magnified  by the artifact of $\sigma_2$ gaps/voids,
    the most noticeable one is around location $100$ of degree
    ($\sigma_1$).  Some articles of scientific experiments have many
    contributing authors, above a threshold typically set by data record
    registration. In particular, the threshold is $100$ by {\tt
      Scopus}. The retained co-authors have at least degree $100$ each,
    many of them may not be co-authors of other articles in the LG
    database and thus the $\sigma_2$ gap at $100$. The co-authors rounded
    off have their $\sigma_2$ counts reduced.
  } %
  \label{fig:sigma-curves}
\end{figure}

We provide in 8 figures and 4 tables the basic information and
empirically analytic data about $6$ epoch-centered collaboration
networks, all drawn from the live literature graph {\tt
  LG-covid19-HOTP}.  The figures and tables are chosen to clarify the
information and packed to fit within the text space limit.  In each
figure/table caption we detail the data description and brief certain
observation(s). Below we elaborate on a few noteworthy phenomena and
valuable findings.

We start with the pairwise comparisons in \cref{tab:eta_intersection-size} among the
networks at the coarsest granularity. To see the comparisons of one
network with the others in the upper or lower triangular, follow the
row and column by the same label (such as MERS) and take the turn at
the diagonal. The intersection set sizes are consistently increasing with
time forward.  However, the agreement profiles are not necessarily
monotonically related. Specifically, among the first five, the
agreement scores of SARS (or Avian flu) with the rest decrease with
time distance; of MERS, peaked with Ebola, and vice versa; of Swine
flu, peaked with MERS. The developing network labeled COVID-19
is more influenced by more recent networks.

The next comparisons are at the cohort level. We see a few phenomena
in \cref{tab:discrepancy-planB-MERS}. The size of cohort (3) is larger
with network more distant from the present, although the network
author size $|V| $ is not monotonically increasing. This correlates
reasonably with the reverse trend in cohort (1). Together they
indicate the lasting and expanding impact of the earlier work by
collective selection and memory of the research community.

We highlight our discovery and understanding of the topological
betweenness encoded by $\sigma_2$, the bi-fork graphlet, which
contributes what lacks with $\sigma_1$ (degree) and $\sigma_4$
(triangle). In \cref{fig:three-snapshot-spectrograms} we point out a
remarkable connection at the persistent cohort (4) between temporal
segmentation and spectral differentiation.  In
\cref{fig:scatter-plots-sigma42} we detail the connection with the
scatter plot of $\sigma_2$-frequencies against $\sigma_4$-frequencies
over the persistent cohort. The triangle is a celebrated motif for the
cluster coefficients, a centrality measure. In fact, the modest bi-fork
plays the critical role of connecting those triangle clusters.  An
author has a triangle connection with every other two co-authors.  All
authors of the same article gain the same amount of triangle count and
degree count as well. In contrast, an author is at the root of a
bi-fork only if the author has more than one articles with different
sets of authors. That is, a bi-fork vertex connects two triangle
clusters. The $\sigma_2$ uniquely encodes such betweenness. 
Furthermore, we take the intersection of all persistent cohorts over
time, and find $103$ authors.  Among them are Malik Peiris, %
the first person to isolate the SARS virus, and Anthony S. Fauci,
well known for his research on HIV and other infectious
diseases. These authors have continued presence and influence in two
decades or longer.  They typically have higher $\sigma_2$ frequencies
than those at the same $\sigma_4$ frequencies.  One may use a similar
approach to measuring temporal-topological betweenness.

We provide in \Cref{fig:sigma-curves} solid evidence of $\sigma_2$
playing another important role in quantifying the grassroots as the
backbone of network connectivity. The data points in all $6$ plots
show unambiguously that while the elite club of hub nodes (with top
degrees) is credited for the small-world phenomenon, the collaboration
network rests largely on the $\sigma_2$ bridge builders.  This finding
implies that a conclusion drawn solely from the $\sigma_1$ degree
information could be much biased to the hub nodes and prone to changes
at the hubs. Our analysis suggests that a network containing larger
mass of nodes with high $\sigma_2$ frequencies is likely more robust
to adversarial changes.  We add that, in its unique roles, the
bi-fork graphlet must co-exist with the triangle graphlet, the count
of bi-forks at each vertex excludes those in any
triangle, see \cref{sec:graphlet-description}.

Our analysis results are readily interpretable. The remarkable
findings reported above are beyond the reach of conventional
approaches. Our method suggests new ways to investigate the dynamics
of author collaboration.

\clearpage
{\small \textbf{Acknowledgements.} This work is partially supported by
  grant 5R01EB028324-02 from the National Institute of Health (NIH),
  USA, and EDULLL 34, co-financed by the European Social Fund (ESF)
  2014-2020.  We are grateful to the reviewer who made numerous
  suggestions on improving the manuscript quality.  We also
  thank Thaleia-M. Passia for helpful comments.}

\bibliographystyle{IEEEtran}

\phantomsection
\label{sec:references}
\addcontentsline{toc}{section}{References}
\balance

\end{document}